\documentclass[letter,11pt]{article}
\usepackage{graphicx}
\usepackage{amsmath, amssymb}
\usepackage[round]{natbib}
\bibpunct{[}{]}{,}{n}{}{,}

\usepackage{subfigure}
\usepackage{booktabs}
\usepackage{xcolor}

\newcommand\eg{\emph{e.g.}}
\newcommand\ie{\emph{i.e.}}

\usepackage{eurosym}
\usepackage{url}

\usepackage{adjustbox}

\usepackage{fancyhdr}
\usepackage[height=8.7in, width=6.5in, top=1.1in, bottom=1.2in,
	headheight=62pt, headsep=0pt, footskip=12pt]{geometry}

\title{\LARGE \bfseries Analyzing In-Game Movements of Soccer Players at Scale}
\author{\Large L\'{a}szl\'{o} Gyarmati, Mohamed Hefeeda \\ \Large Qatar Computing Research Institute, HBKU \\ \{lgyarmati,mhefeeda\}@qf.org.qa}
\date{}

\begin{document}

\maketitle

\begin{abstract}
	It is challenging to get access to datasets related to the physical performance of soccer players. The teams consider such information highly confidential, especially if it covers in-game performance.Hence, most of the analysis and evaluation of the players' performance do not contain much information on the physical aspect of the game, creating a blindspot in performance analysis.
	We propose a novel method to solve this issue by deriving movement characteristics of soccer players. We use event-based datasets from data provider companies covering 50+ soccer leagues allowing us to analyze the movement profiles of potentially tens of thousands of players without any major investment. Our methodology does not require expensive, dedicated player tracking system deployed in the stadium. We also compute the similarity of the players based on their movement characteristics and as such identify potential candidates who may be able to replace a given player. Finally, we quantify the uniqueness and consistency of players in terms of their in-game movements. Our study is the first of its kind that focuses on the movements of soccer players at scale, while it derives novel, actionable insights for the soccer industry from event-based datasets.
\end{abstract}

\section{Introduction}
Quantitative performance analysis in sports has become mainstream in the last decade. The focus of the analyses is shifting towards more sport-specific metrics due to novel technologies. These systems measure the movements of the players and the events happening during trainings and games. This allows for a more detailed evaluation of professional athletes with implications on areas such as opponent scouting, planning of training sessions, or player scouting.

Previous works that analyze soccer-related logs focus on the game-related performance of the players and teams. Vast majority of these methodologies concentrate on descriptive statistics that capture some part of the players' strategy. For example, in case of soccer, the average number of shots, goals, fouls, passes are derived both for the teams and for the players~\cite{anderson2013numbers,duch2010quantifying}. Other works identify and analyze the outcome of the strategies that teams apply~\cite{pena2012network,narizuka2013statistical,lucey2013assessing,gyarmati2014,gyarmati2015,Wang2015,luceyquality}. However, the physical performance and in particular the movements of players has not received detailed attention yet. 

It is challenging to get access to datasets related to the physical performance of soccer players. The teams consider such information highly confidential, especially if it covers in-game performance. Despite the fact that numerous teams deployed player tracking systems in their stadiums, datasets of this nature are not available for research or for public usage. It is nearly impossible to have quantitative information on the physical performance of all the teams of a competition. Hence, most of the analysis and evaluation of the players' performance do not contain much information on the physical aspect of the game, creating a blindspot in performance analysis.

We propose a novel method to solve this issue by deriving movement characteristics of soccer players. We use event-based datasets from data provider companies covering 50+ soccer leagues allowing us to analyze the movement profiles of potentially tens of thousands of players without any major investment. Our methodology does not require expensive, dedicated player tracking system deployed in the stadium. Instead, if the game is broadcasted, our methodology can be used. As a consequence, our technique does not require the consent of the involved teams yet it can provide insights on the physical performance of many players in different teams. 

%

The main contribution of our work is threefold:
\begin{enumerate}
	\item we propose a methodology to extract the movement characteristics of the players,
	\item we compute the similarity of the players and as such identify potential candidates who may be able to replace a given player,
	\item we quantify the uniqueness and consistency of players in terms of their in-game movements.
\end{enumerate}
To the best of our knowledge, our study is the first of its kind that focuses on the movements of soccer players at scale, while it derives novel, actionable insights for the soccer industry from event-based datasets.

\section{The Challenge of Profiling Movements}
As we noted already, it is not straightforward how to quantify the movements of soccer players at scale. The core of this problem lies in the properties of the potential datasets that may be used for the process. There exist three main data acquisition methodologies applied in the soccer industry: event-based, tracking, and wearable sensors. We briefly describe each of them focusing on their properties relevant to the analysis of the players' movements.

First, event-based datasets annotate the most important, ball related events of a game. The method involves human operators who code the games based on a corresponding video feed of the game. Although data providers apply quality assurance techniques\footnote{\eg, multiple operators annotate the game and the final data feed is a result of majority voting.}, this technique is prone to human errors. Despite of this, the datasets are widely used in the media to enhance the fan experience during the game. On the other hand, the data feed is near real-time and the data production does not need any dedicated system to be deployed at the stadiums.

Second, tracking datasets contain fine-grain details on the movement of players and of the ball throughout the game. This data is generated based on video feeds of dedicated, precisely positioned cameras. Optical tracking algorithms extract the trajectories from the video; however, there are scenarios (\eg, collision of players) where the supervision of human operators are needed. A recent study revealed that discrepancies exist among different tracking systems, \eg, the trajectories of a player may be of within several meters~\cite{IanGrahamTalkFCBSymp}. A major drawback of this technique is that it involves the deployment of a system in the stadium. As such, the consent of the home team is mandatory for such data collection. Anyone who intends to analyze the movements of the players of a competition should get the consent of all teams (and usually of the league too). This is a major obstacle for physical performance analysis at scale.

Third, wearable sensor devices collect detailed datasets on the movement of the players~\cite{gpsports,statsports,catapult}. The sensors of these devices capture the movement, accelerations, and rotation of the players, among others. The in-game application of this technique was authorized by a recent decision of FIFA, the governing body of international soccer~\cite{FIFASensor}. As of July 2015, players are allowed to wear sensors during official games. However, recent research studies report discrepancies related to the precision and consistency of these devices and such data should be used with precaution~\cite{buchheit2014integrating}. There is a more crucial practical issue with this technique: the dataset holds information only on the players of a single team, details on the ball and the opponent players are missing. This prevents any comparisons of players from multiple teams, and tactical analysis of players.

The review on the available data acquisition techniques reveals the difficulty of any study focusing on the quantitative evaluation of the players' movements on scale. As we show in this paper, event-based datasets can be used to address this problem and provide insights on player movements. We introduce our methodology in the next section.




\section{Methodology}
In this section, we introduce our methodology used to extract the movements of players and then to create their movement characteristics. Our final goal is to quantify the similarities of players based on their movement characteristics, \ie, the movements they apply during a season. We use an event-based dataset throughout our analyses that we describe next.

\subsection{Dataset}
We use an event-based dataset generated by Opta~\cite{opta} covering the 2012/13 season of La Liga (\ie, the first division soccer league of Spain). The dataset contains all major events of a soccer game including passes, shots, dribbles, and tackles. For example, the dataset has more than 300,000 passes and nearly 10,000 shots.
The dataset contains the time and location of these events along with the identity of the involved players. 
Hence, it is possible to derive a coarse grain time-series of the movements of the players~\cite{Gyarmati2015Porto}. We note that the precision of the time annotation of the dataset is one second.

\subsection{Movement vector extraction}
\begin{figure}[tb]
\centering
\subfigure[Xavi]{\adjincludegraphics[clip=true, trim=1cm 2.5cm 1cm 3cm,width=5.4cm]{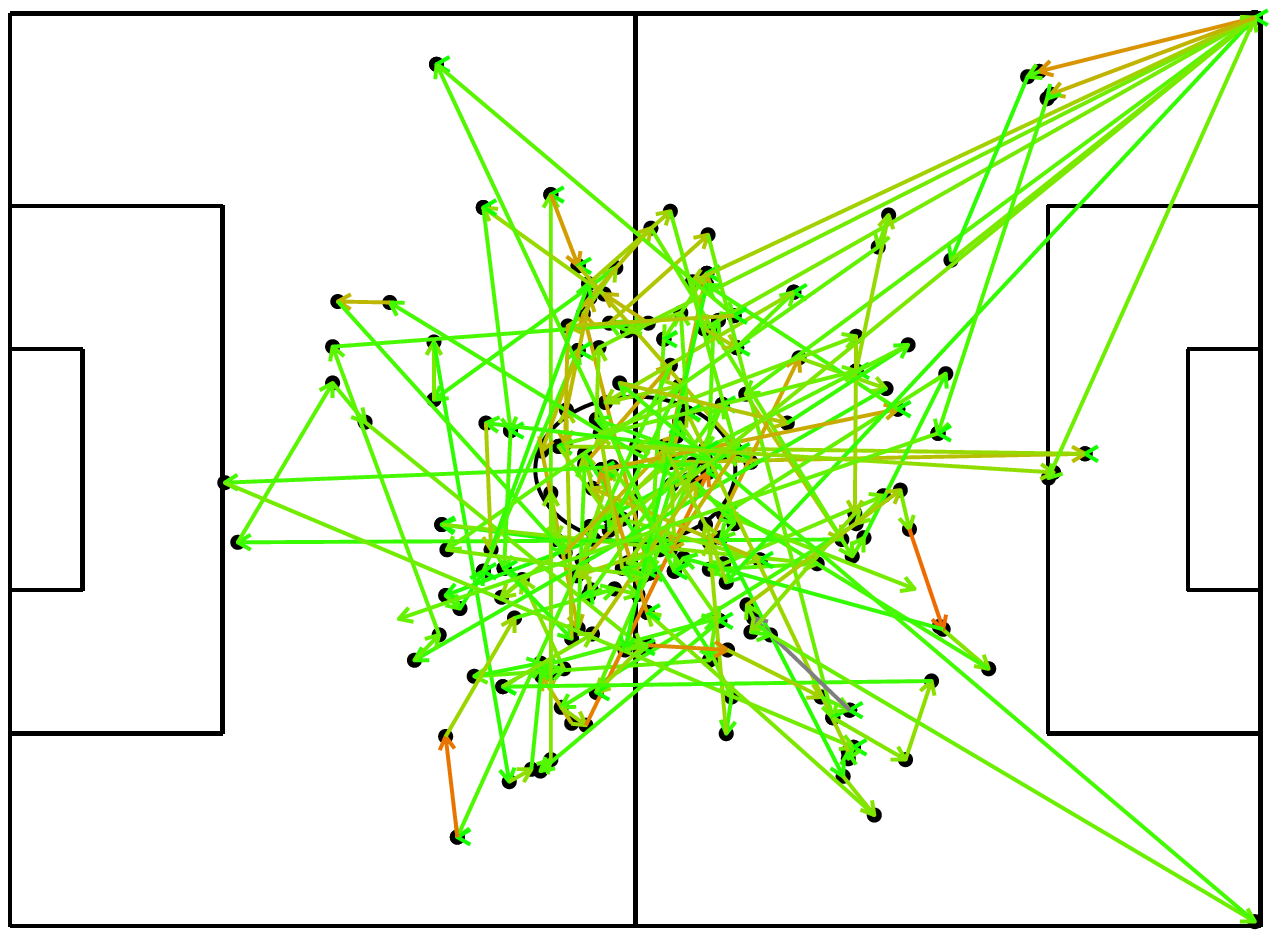}\label{fig:movement_vectors_xavi}}
\subfigure[Messi]{\adjincludegraphics[clip=true, trim=1cm 2.5cm 1cm 3cm,width=5.4cm]{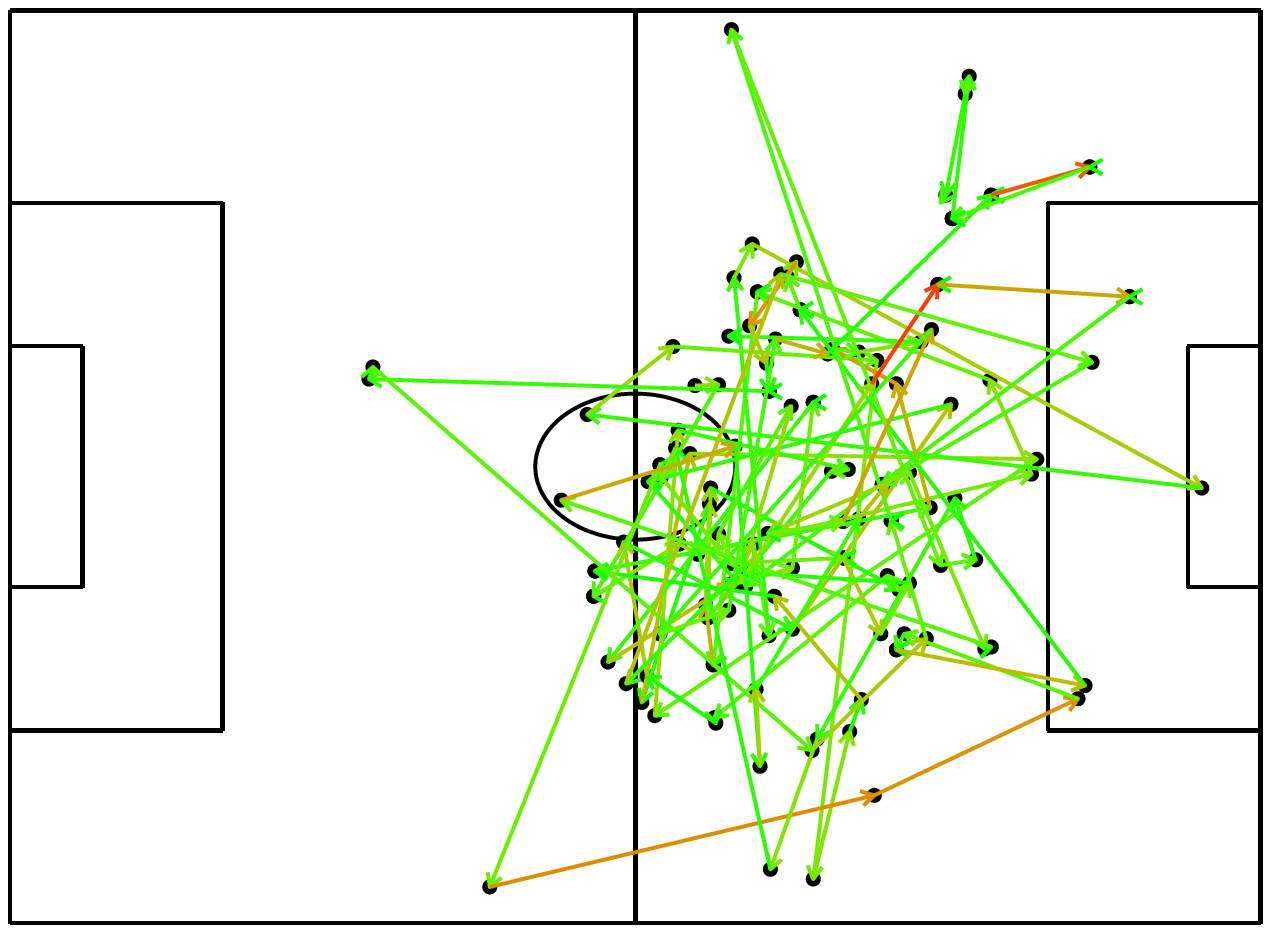}}
\subfigure[Ronaldo]{\adjincludegraphics[clip=true, trim=1cm 2.5cm 1cm 3cm,width=5.4cm]{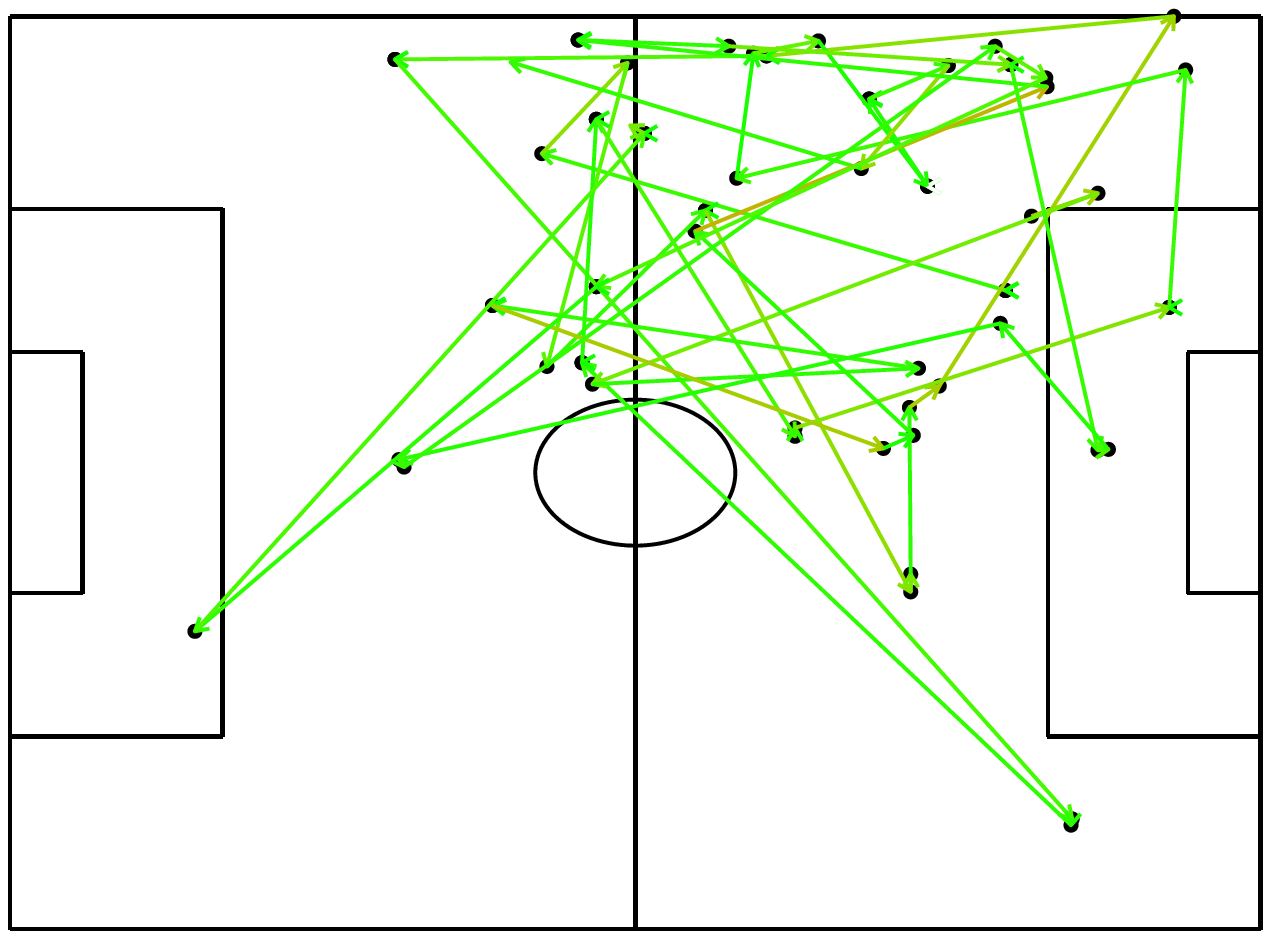}}
\caption{Players' movement vectors in a game derived from an event-based dataset. The color of the arrows correspond to the speed of the movements (green---slow, red---fast). The teams are attacking from left to right.}
\label{fig:movement_vectors}
\end{figure}

We describe each movement as a vector of seven: $(x_1,y_1,x_2,y_2,T,s,b)$, where the movement starts at time $T$ at location $(x_1,y_1)$, ends at location $(x_2,y_2)$, with speed $s$, while $b$ denotes ball possession (\ie, whether the player had the ball or not). In total, we derive 660,848 movement vectors of 542 players for the analyzed season. The players have diverse movements over the season---both in terms of their numbers and properties: the mean number of movements per player is 1,219 (could be as high as 4,998), while the mean length of the movements is 19.4 meters (up to 100 meters). We illustrate the derived movements of three players in Figure~\ref{fig:movement_vectors} given a single game. The figures show the area where the players tend to move and also reveal their role in the team. For example, Xavi was active mainly in the middle of the field and had some high intensity movements (arrows with red color). Messi leaned towards the right side of the field and penetrated the box of the opponent five times (arrows pointing into the box). Cristiano Ronaldo on the other hand was moving on the left side, his movements covered larger distances. This is the first step of our methodology: extracting the movement vectors of the players. We note that the event-based dataset we use is sparse in terms of the position of the players, \ie, the physical location of a player is only recorded when the player was involved in some ball-related event\footnote{This is a consequence of the data acquisition process: the games are annotated based on the television broadcast that focuses on the ball all the time.}. As such, the elapsed time between two events of a player can be as low as couple of seconds but it can reach several minutes too. 

During the creation of the movement vectors, we take into account that the size of the soccer fields are not necessarily identical. An interesting property of the rules of soccer is that the sizes of the field are not fixed, there is some room to design a soccer pitch even in case of international matches. According to the first law of the game, the length of the pitch shall be between 100 and 110 meters, while the width between 64 and 75 meters~\cite{FIFArules}. There is an ongoing standardization effort, most of the newly constructed stadiums have a pitch with a size of 105x68m\cite{uefa_pitch}. Spain is not an exception to this extent, where the dimensions of Elche's stadium are 108x70m while the field is 100x65m in case of Rayo Vallecano~\cite{spain_stadium_sizes}. 

Another data preparation technique we use is related to handling the passes in the dataset. For passes, we have a complete datapoint for the initiator of the pass (\ie, timestamp and location), however, at the receiving end, the dataset does not contain a timestamp. To overcome this issues, and to increase the wealth of the extracted time-series, we use the timestamp of the previous event, \ie, the initiation of the pass. This is the best method for estimation as described in~\cite{Gyarmati2015Porto}.



\subsection{Movement characteristics construction}
\begin{figure}[tb]
\centering
\subfigure[All movements]{\adjincludegraphics[clip=true, trim=1cm 2.5cm 1cm 3cm,width=5.4cm]{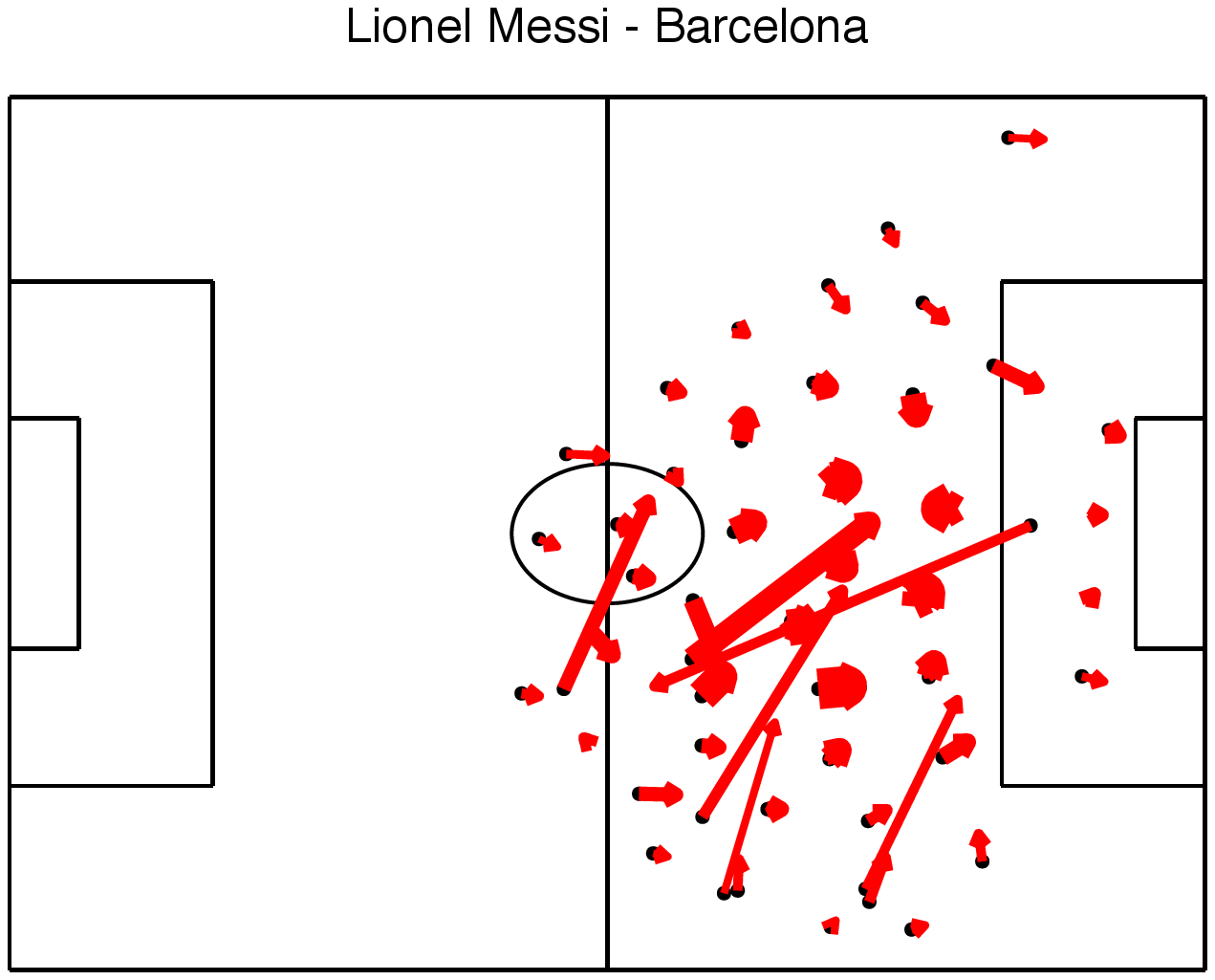}\label{fig:movement_characteristics_all}}
\subfigure[Movements with ball]{\adjincludegraphics[clip=true, trim=1cm 2.5cm 1cm 3cm,width=5.4cm]{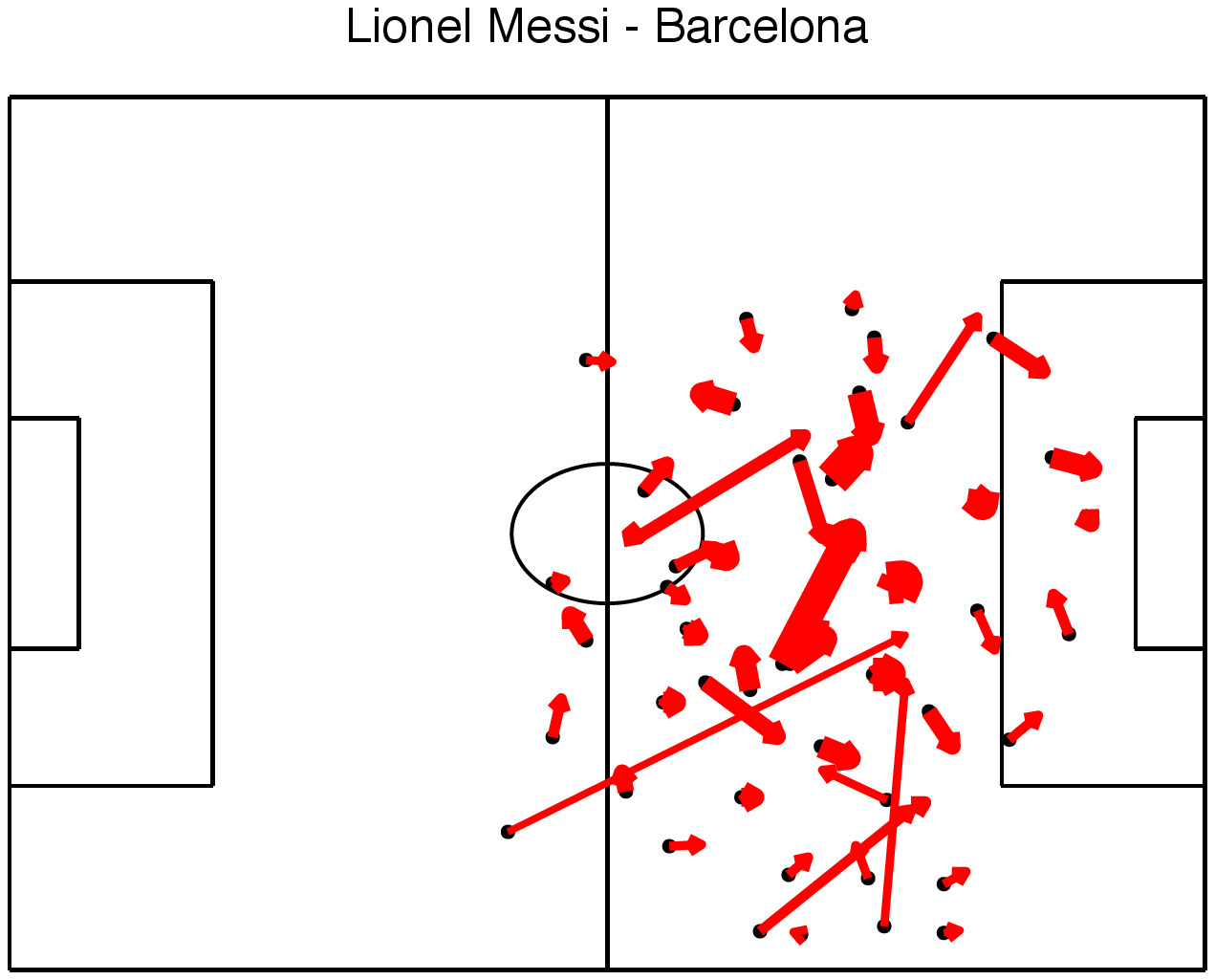}\label{fig:movement_characteristics_ball}}
\subfigure[High speed movements ($\geq 14km/h$)]{\adjincludegraphics[clip=true, trim=1cm 2.5cm 1cm 3cm,width=5.4cm]{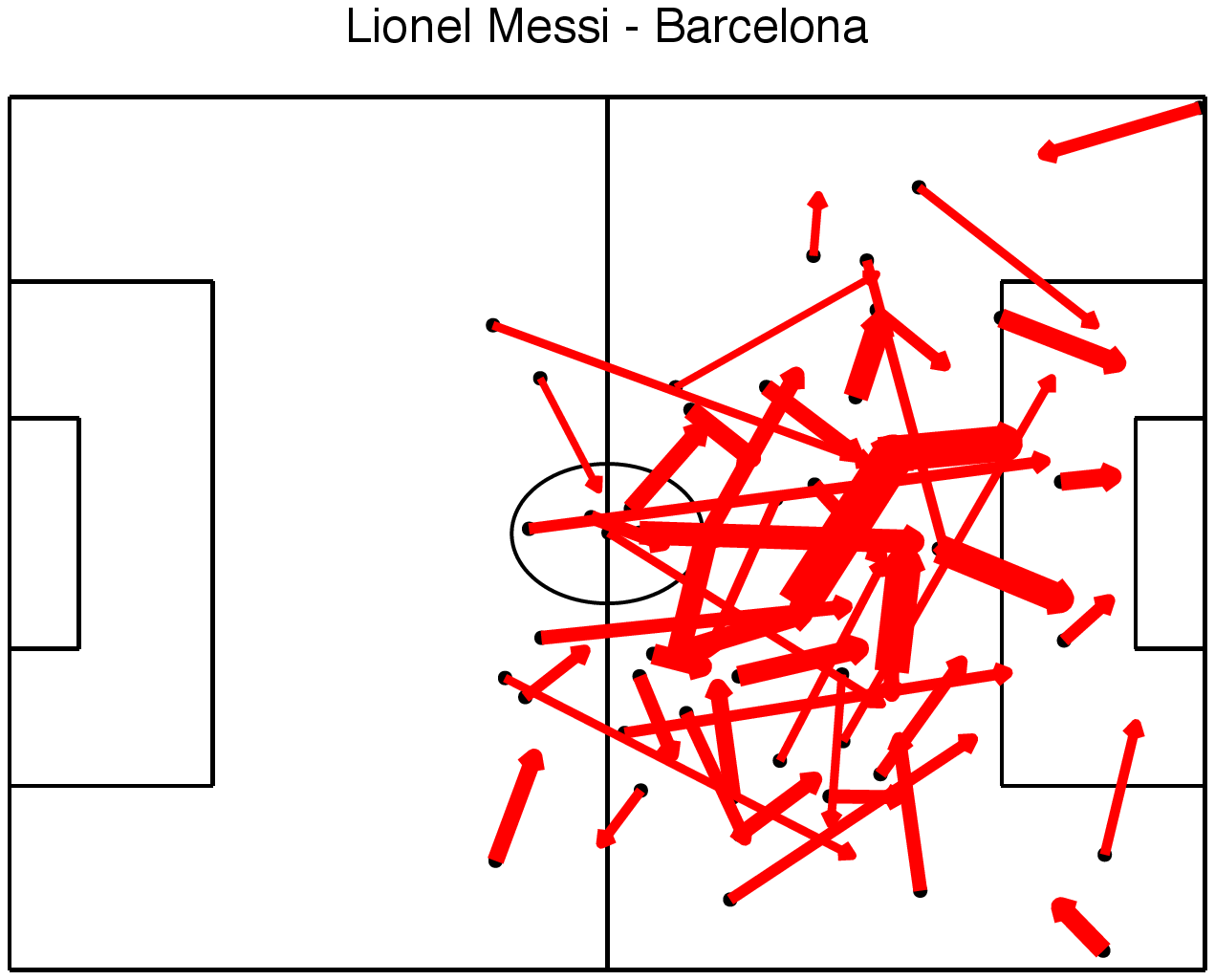}\label{fig:movement_characteristics_run}}
\caption{The most important 50 movement directions of Messi throughout the season. The boldness of the arrows correspond to the frequency the movement was used. Depending on the set of movements we consider, the methodology reveals diverse insights on the player's physical performance.}
\label{fig:movement_characteristics}
\end{figure}

\begin{figure}[h]
\centering
\subfigure[]{\adjincludegraphics[clip=true, trim=1cm 2.5cm 1cm 3cm,width=5.4cm]{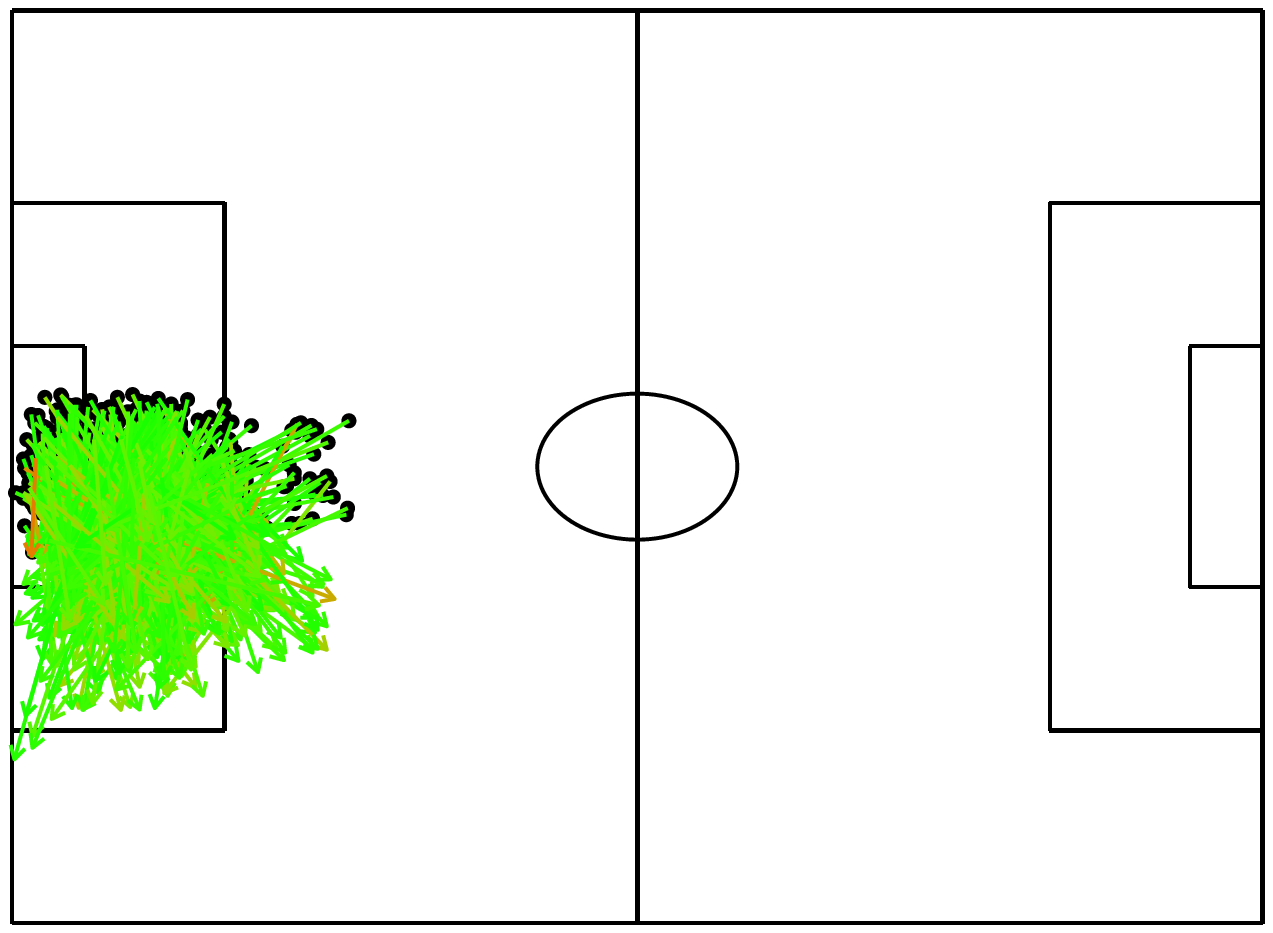}}
\subfigure[]{\adjincludegraphics[clip=true, trim=1cm 2.5cm 1cm 3cm,width=5.4cm]{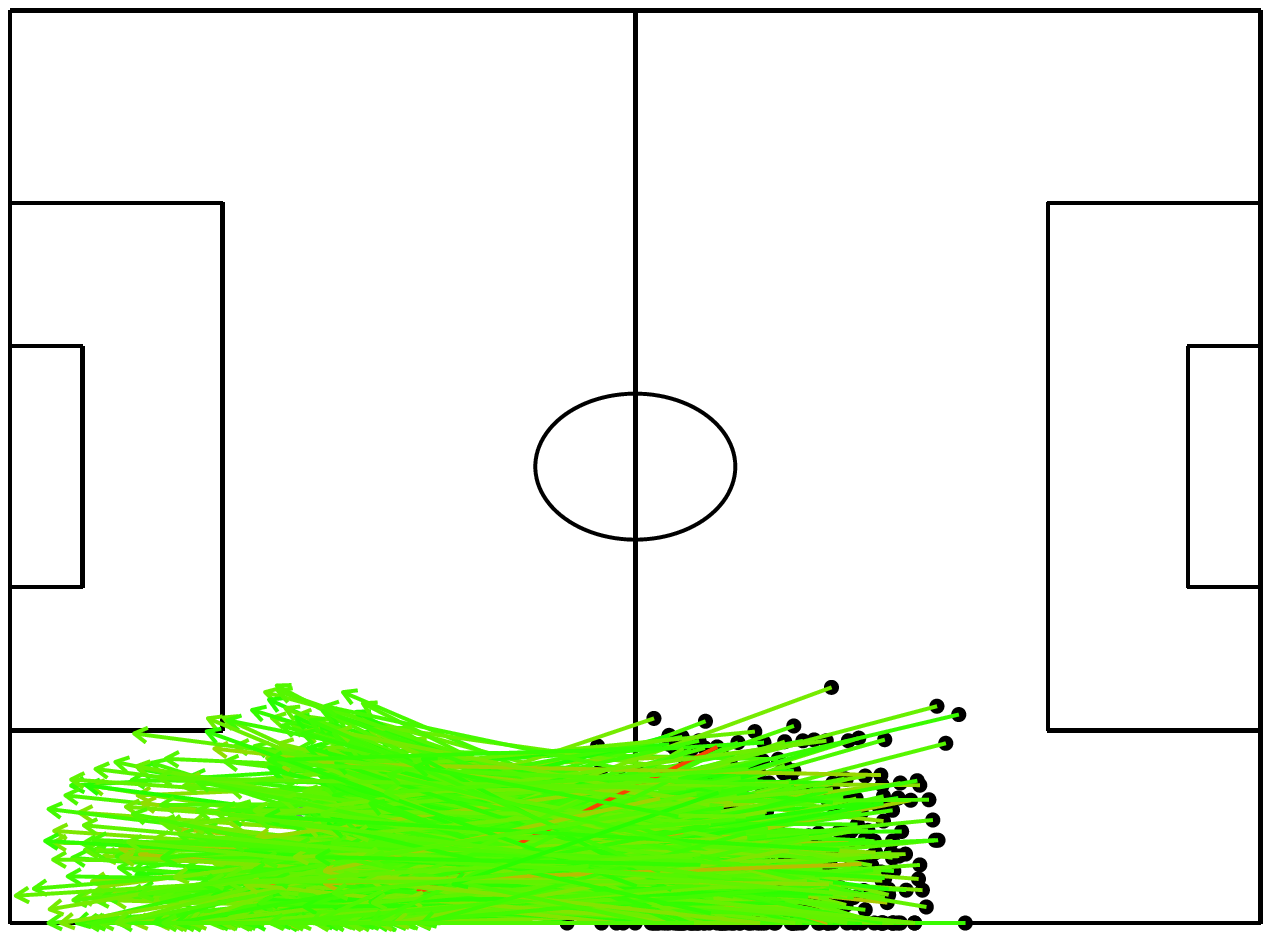}}
\subfigure[]{\adjincludegraphics[clip=true, trim=1cm 2.5cm 1cm 3cm,width=5.4cm]{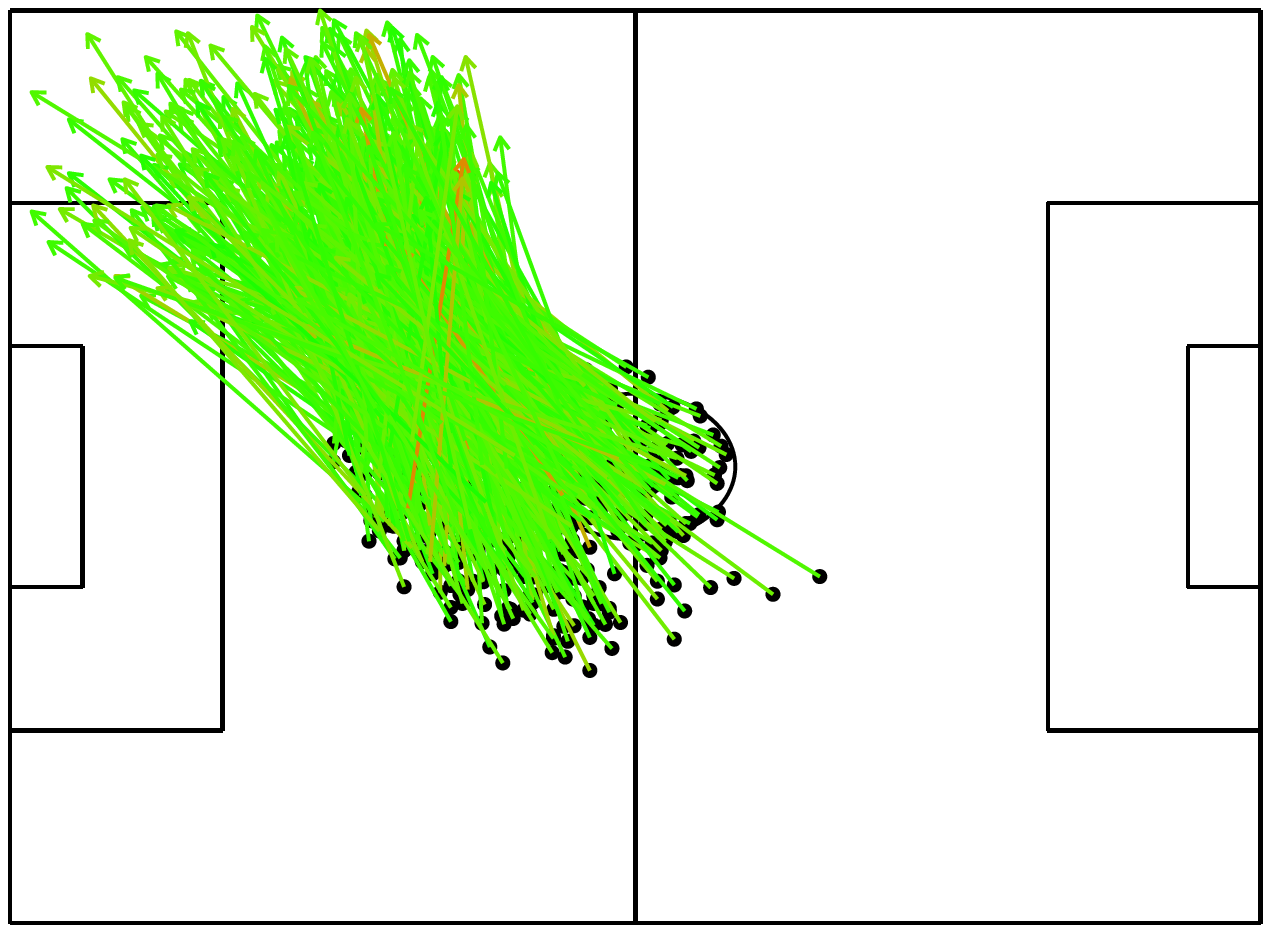}}
\subfigure[]{\adjincludegraphics[clip=true, trim=1cm 2.5cm 1cm 3cm,width=5.4cm]{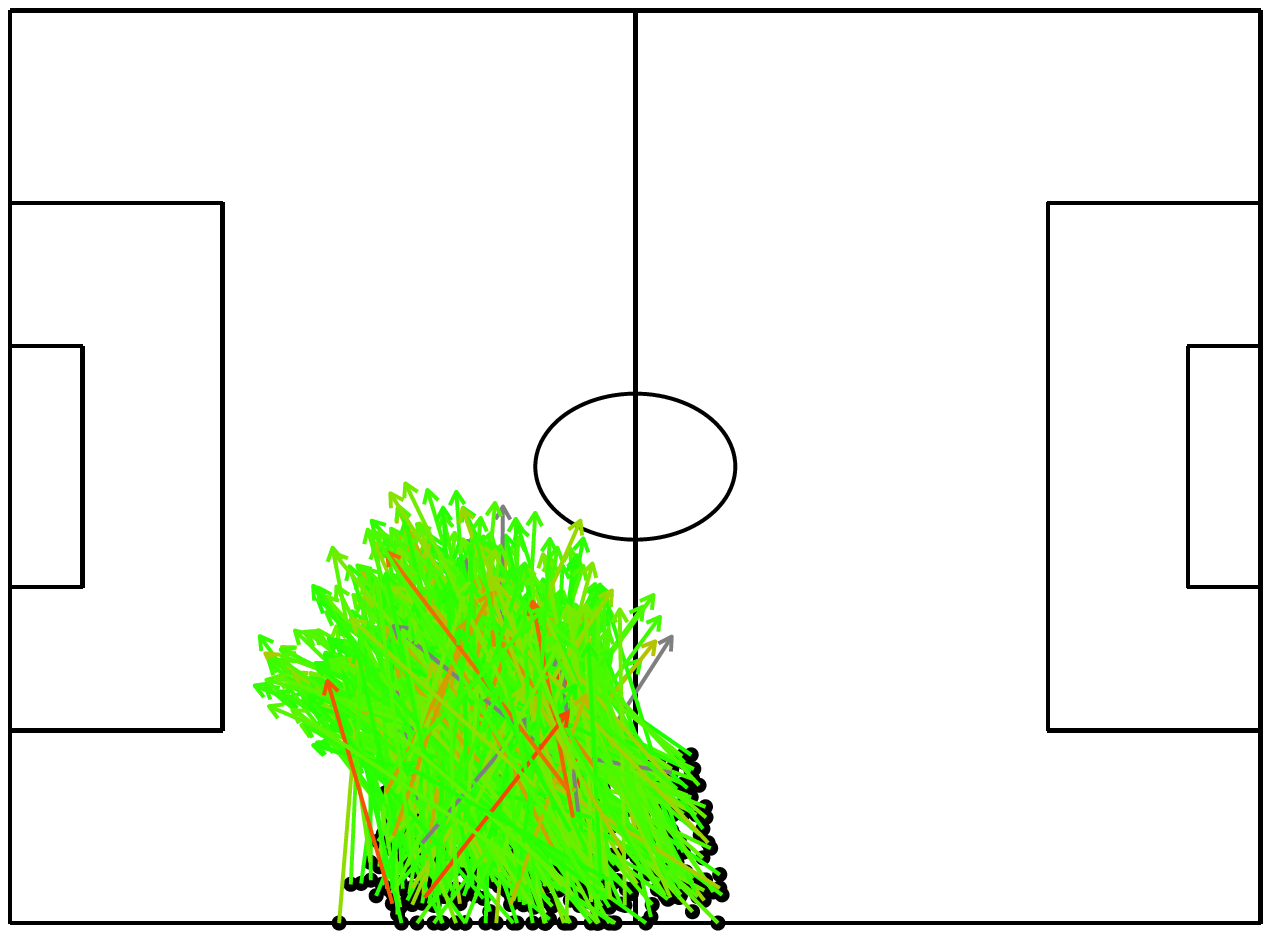}}
\subfigure[]{\adjincludegraphics[clip=true, trim=1cm 2.5cm 1cm 3cm,width=5.4cm]{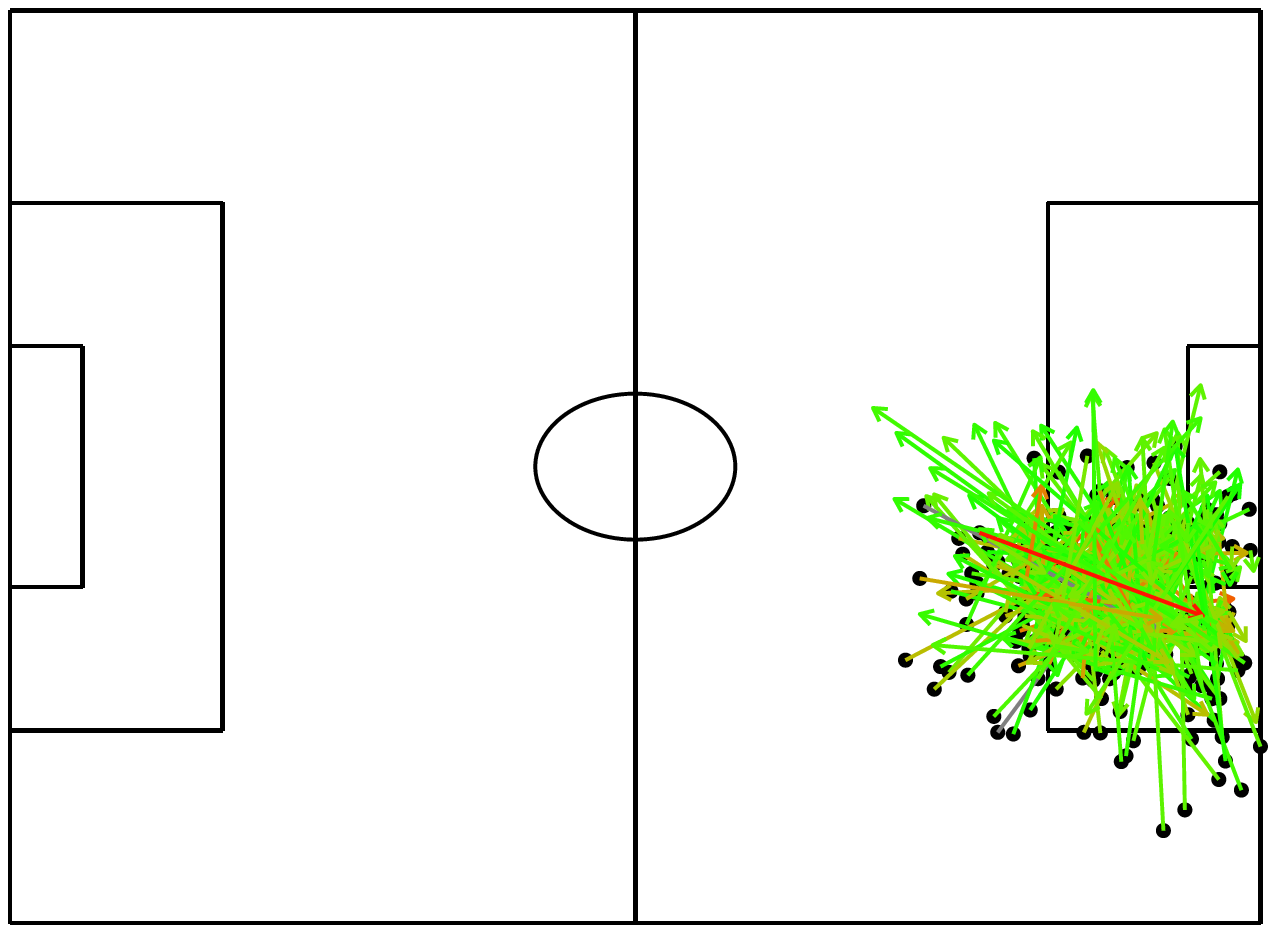}}
\subfigure[]{\adjincludegraphics[clip=true, trim=1cm 2.5cm 1cm 3cm,width=5.4cm]{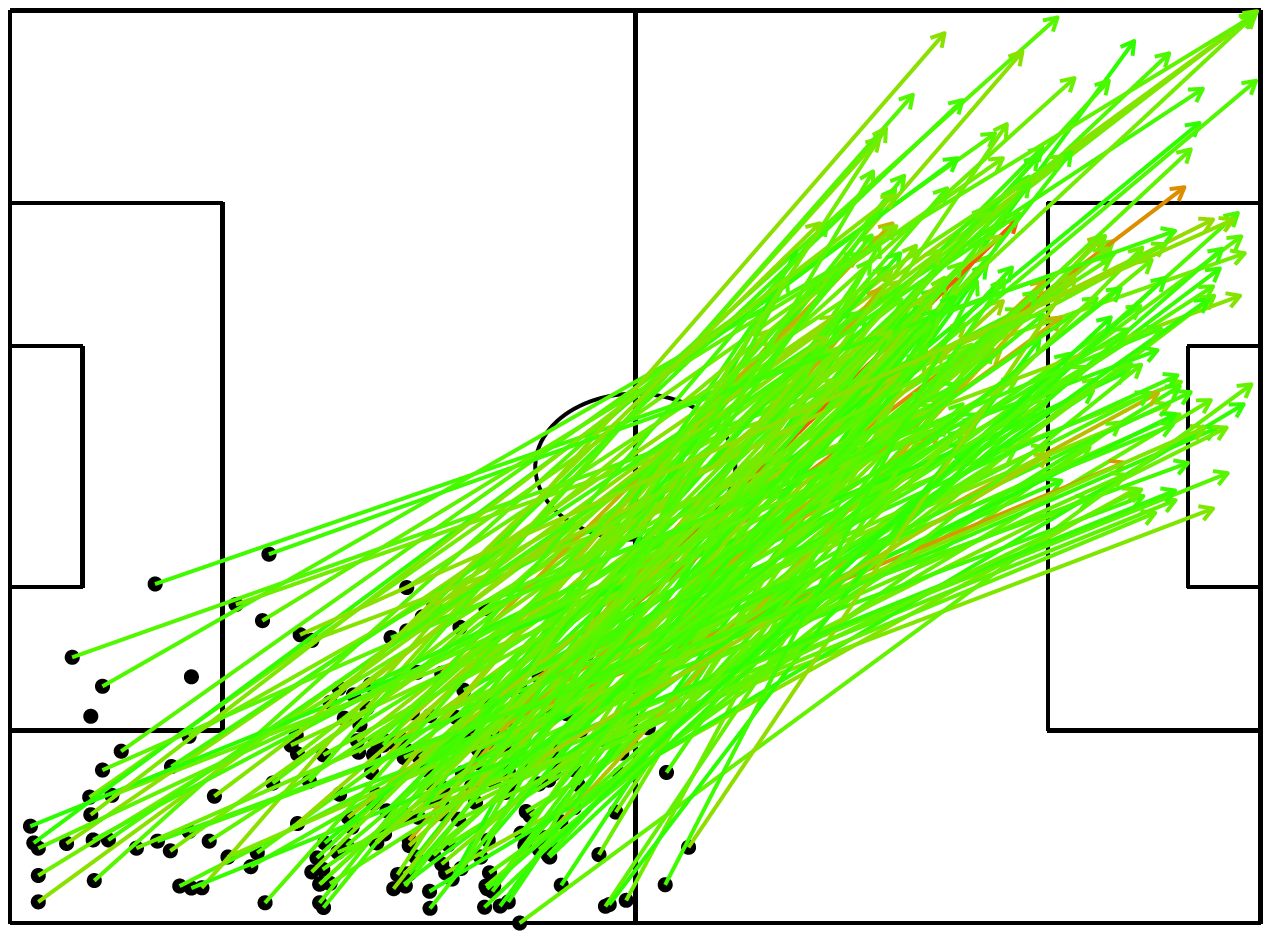}}
\caption{The coverage of some of the feature vectors (out of 200). The area of the clusters are not identical, as such, most important parts of the field are handled with more fine-grain details.}
\label{fig:feature_vector_spread}
\end{figure}

Our goal is to derive movement characteristics that enable us comparing the players’ performance and analyzing the stability of a player’s role and fitness across a season. Players have diverse number of vectors, to handle this, we apply the following methodology. First, we derive the most relevant $K$ movement vectors of the competition using all the vectors of all the players. We determine these features using the mini-batch $K$-means clustering algorithm~\cite{sculley2010web}: the centroid movement represents the vectors belonging to a specific cluster. We apply this method instead of creating a grid for the locations, to have smooth, balanced clusters (instead of having high skew among the number of members in the clusters in case of the grid scheme). Throughout this paper, we use $K=200$. Second, in case of each movement of a player, we determine the cluster it belongs to, \ie, we compute the most similar feature vector. In the Appendix, we show examples of the coverage of some of the feature vectors, \ie, which movement vectors are belonging to a given feature vector (Figure~\ref{fig:feature_vector_spread}). Third, we aggregate the number of times a player applied each feature vector, which creates a frequency vector of the features. Finally, we normalize the frequencies with the total number of movements a player has. As a result of the normalization, we have the movement characteristic of a player. As an example, we present the top 50 movement directions of Messi in Figure~\ref{fig:movement_characteristics_all}. The figure reveals that Messi tends to have short movements in the final third of the pitch, while his mid/long range movements are starting from the right side of the field.

We are able to focus on specific movements of the players by applying filters on the initial set of movement vectors. One such filter is ball possession. It is a crucial insight how a player moves while having the ball. To focus on this, we determine all the combinations of the events where a player has the ball (\ie, both the starting and ending events of the movement should involve ball possession). Events like recovering the ball, intercepting the ball, and pass reception mark the beginning of movements where the player has possession of the ball, i.e., the player is moving together with the ball. On the contrary, if the first event of a movement is making a pass, the player does not have the ball for the given movement vector. After the filtering, we construct the feature vectors and the characteristic vectors of the players. We present the most important with ball movement traits of Messi in Figure~\ref{fig:movement_characteristics_ball}. There are six major routes Messi takes when he has the ball (shown with thick arrows). All of these movements are located centrally at the beginning of the final third.

Another aspect is the speed of the movements: fast movements are generally collocated with important events of the game. We apply a threshold on the speed of the movements, \ie, the movements ought to be at least 14km/h. This is inline with the categories widely used in the soccer industry~\cite{Valter2006, Bangsbo1991}. Figure~\ref{fig:movement_characteristics_run} presents the high-speed movements of Messi: not only his favorite traits on the midfield are revealed but also the tendencies how he approaches and enters the box of the opponent.


\subsection{Uniqueness and consistency}
One of the main applications of the movement characteristics is finding similar players who may be able to replace a given player. We identify candidates to replace some players later on this paper, some similarities may be surprising on first sight. However, this is not the only insight we can gain from the profiles. The movement characteristics of the players enable us to quantify two additional decisive qualities of the players: uniqueness and consistency. For this, we use the cosine similarity to measure the distance between two players. Uniqueness measures how hard it is to find a player that has similar movements. We use the movement characteristics derived over the entire season, and for each player we determine the $M$ most similar ones ($M=5$ in our evaluations). This is done by identifying the players with the smallest distance from the particular player (\ie, players with minimal cosine distances). Let $d_{ij}=D(c_i,c_j)$ denote the cosine distance between player $i$ and $j$, where $c_i$ denotes the movement characteristic of player $i$. We compute the uniqueness of player $i$ as:
\begin{equation}
	U_i = \sum_{j=1}^{M}d_{ij}
\end{equation}
The range of this metric is $(0,M)$, the higher this value is the more unique a player is. The uniqueness metric can be generalized for game specific movements: in this case the distances are measured between the movement characteristics of individual games, not the entire season.

It is preferable if a player applies movements that are hard to reproduce. However, it is equally important to have consistent performance throughout the season. We evaluate this using the game-wise movement characteristics of the players. Let $t=1,\dots,N$ denote the games a player was involved in, while $c_i^k$ denote the movement characteristic of player $i$ for game $k$. The consistency of player $i$ for game $k$ is defined as the average pairwise distance of its movement characteristics:
\begin{equation}
	C_i^k = \frac{1}{N}\sum_{t=1}^{N}D(c_i^k,c_i^t)
\end{equation}
The range of consistency is $(0,1)$. If the consistency metric $C$ is small, the player applies similar movements across the whole season, \ie, it is expected to see the same kind of movements throughout the season.



\section{Empirical Results and Insights}
We next apply the proposed methodology on the presented dataset covering the events of the 2012/13 season of the Spanish first division soccer league. We first focus on identifying similar players, afterwards we study the uniqueness and consistency of the players. Finally, we highlight an additional area in which the methodology is able to derive new insights: movements related to creating chances.


\subsection{Similarity}
\begin{figure}[tb]
\centering
\subfigure[Cristiano Ronaldo (Real Madrid)]{\adjincludegraphics[clip=true, trim=1cm 2.5cm 1cm 3cm,width=7.5cm]{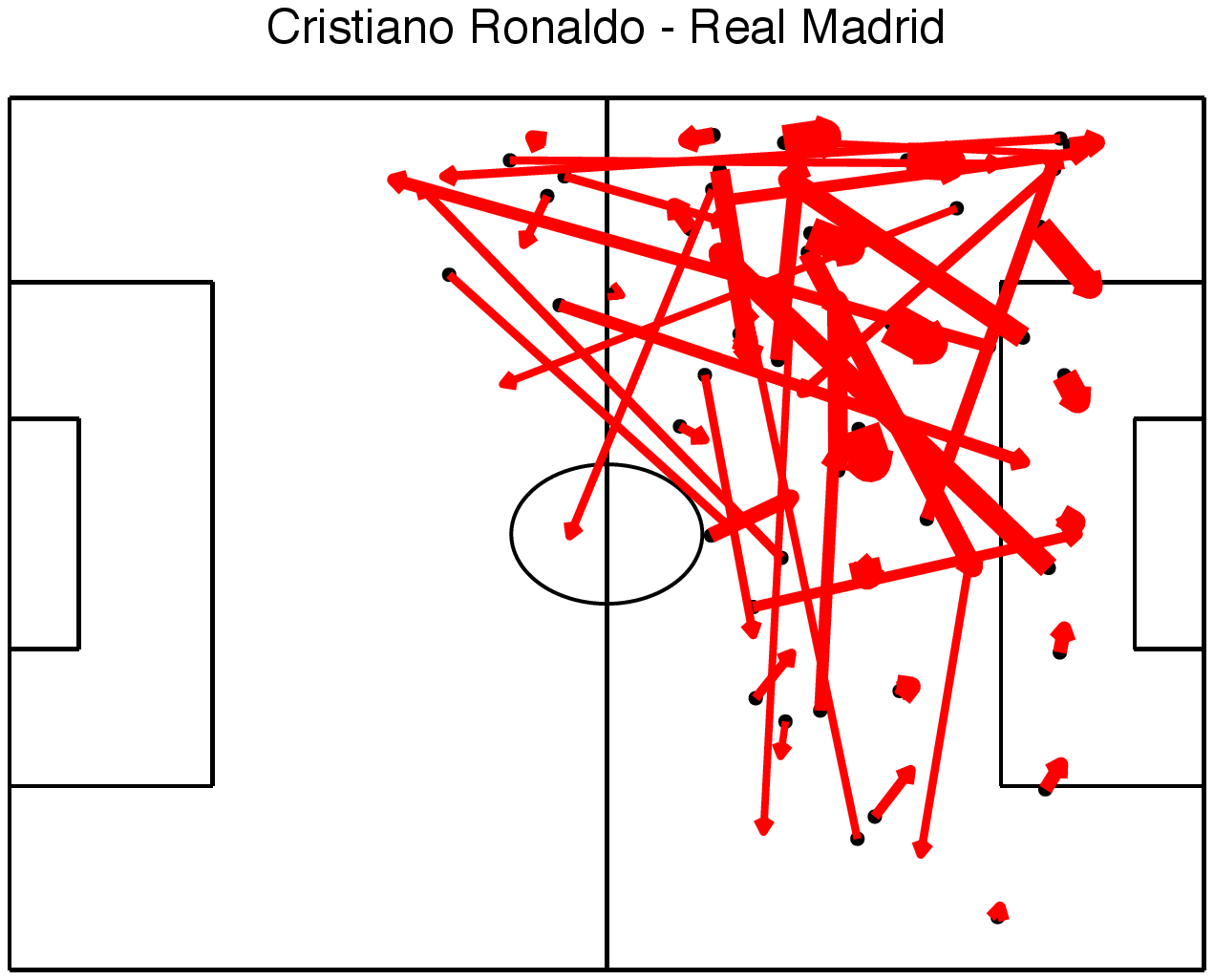}}
\subfigure[Ruben Castro (Real Betis)]{\adjincludegraphics[clip=true, trim=1cm 2.5cm 1cm 3cm,width=7.5cm]{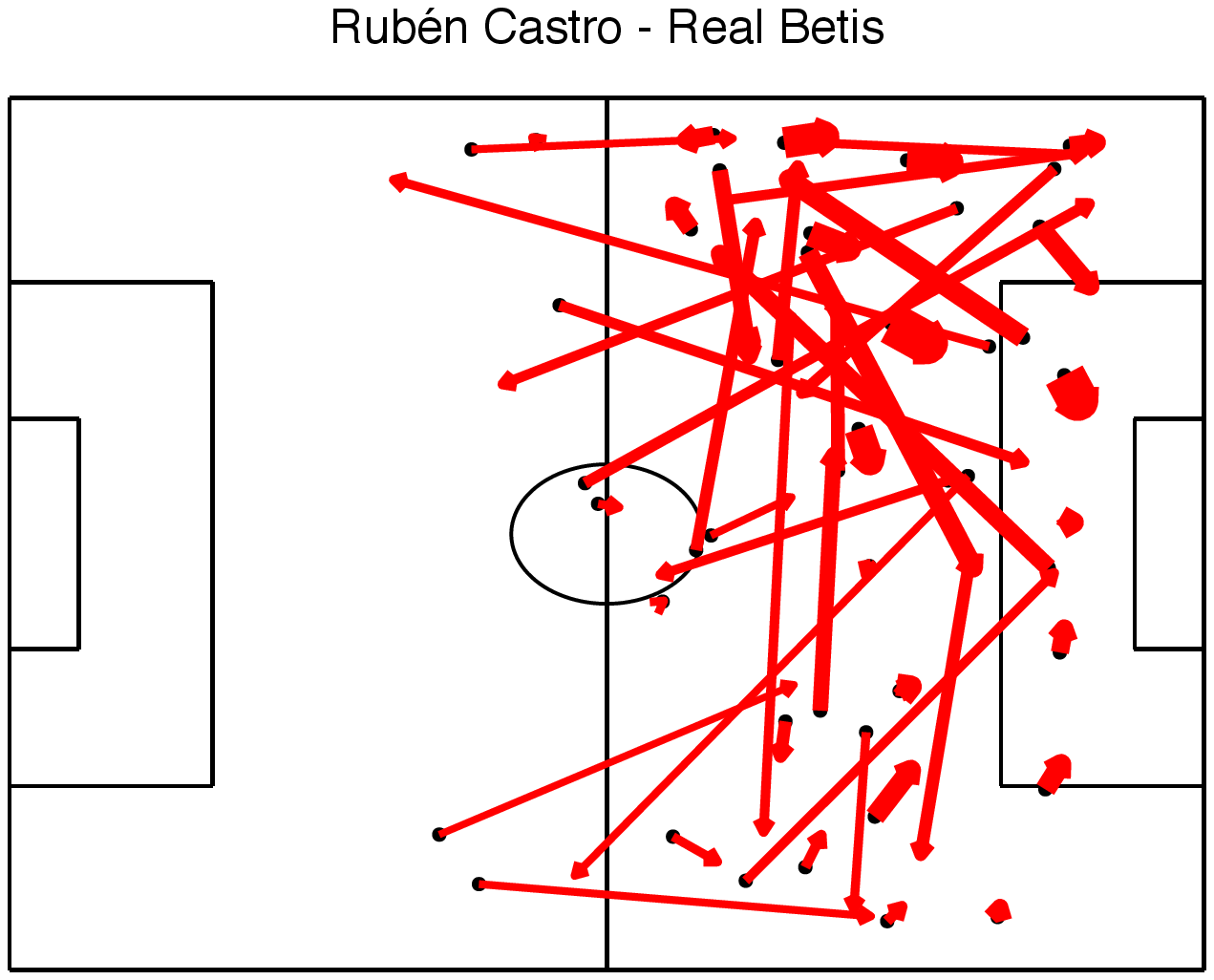}}
\caption{Top-50 movement features of Cristiano Ronaldo and his most similar counterpart, Ruben Castro. There is a remarkable similarity between the two players, while having two orders of magnitude difference in their market values.}
\label{fig:player_similarity}
\end{figure}

We determine the similarity of the players based on their movement characteristics derived using all the movements the players had during the season. In Table~\ref{tab:most_similar_players} we focus on Messi, Cristiano Ronaldo, and Xavi by presenting their five most similar counterparts. The list of similar players may be considered as a shortlist of candidates who are potentially able to replace the given player---at least based on their in-game movements. The table contains the distance of the players from each other.  We show the market value of the players in the table as well as a reference. The market values of the players are as of the end of the season (\ie, June 2013) based on the estimations of Tranfermarkt~\cite{transfermarkt}. Some of the results are straightforward like the case of Messi and Saviola, or Xavi and Thiago. However, it is interesting to see the list of Cristiano Ronaldo. Our methodology reveals that the most similar player of Cristiano Ronaldo was Ruben Castro (of Real Betis). We illustrate the similarity of the movements of Ronaldo and Castro in Figure~\ref{fig:player_similarity}. The figure shows the feature vectors the players used and to what extent. There is a remarkable similarity between the players despite the fact that Castro is not as highly rated as Ronaldo, and there is a huge discrepancy in their market values: 100M compared to 4.5M. This example highlights the most important benefit of our scheme: we are able to identify players who are having the same kind of movements as their more famous reference, however, for only a fraction of the price.

\begin{table}[tb]
	\scriptsize
	\centering
\begin{tabular}{clcc}
\toprule
 \# similar &  player & distance & market value (\euro)  \\
\midrule
\midrule                 
 &  Lionel Messi - FC Barcelona &  &  120M   \\
\midrule
1 & Javier Saviola - Málaga & 0.155 & 3M \\
2 & Radamel Falcao - Atlético Madrid & 0.158 & 60M \\
3 & Diego Buonanotte - Granada CF & 0.174 & 2M \\
4 & Obafemi Martins - Levante & 0.180 & 3.5M \\
5 & Enrique De Lucas - Celta de Vigo & 0.193 & 0.5M \\
\midrule
\midrule
& Cristiano Ronaldo - Real Madrid &  & 100M \\
\midrule
1 & Rubén Castro - Real Betis & 0.079 & 4.5M \\
2 & Antoine Griezmann - Real Sociedad & 0.089 & 15M \\
3 & Helder Postiga - Real Zaragoza & 0.125 & 5M \\
4 & Jorge Molina - Real Betis & 0.127 & 3.5M \\
5 & Jonathan Viera Ramos - Valencia & 0.132 & 3M \\
\midrule
\midrule
 & Xavi Hernández - FC Barcelona &  & 15M \\
\midrule
1 & Thiago Alcántara - FC Barcelona & 0.069 & 22M \\
2 & Sami Khedira - Real Madrid & 0.109 & 22M \\
3 & Luka Modric - Real Madrid & 0.113 & 35M \\
4 & Ignacio Insa - Celta de Vigo & 0.116 & 0.9M \\
5 & Daniel Parejo - Valencia & 0.122 & 10M \\
\bottomrule
\end{tabular}
\caption{The top five most similar players and their market values in case of Messi, Cristiano Ronaldo, and Xavi.}
\label{tab:most_similar_players}
\end{table}

\subsection{Uniqueness}
We next focus on the uniqueness of the players, \ie, how hard it is to find a player who is able to execute the same in-game movements throughout a season. In case of uniqueness, we only consider players who had at least $500$ movements across the season to avoid discrepancies due to players with small participation in the season. We show the ten most unique players of the competition in Table~\ref{tab:uniqueness}. As a reference, we also include the number of movements the players had in the season. The majority of these players are defenders who were playing significant portions of the season on different sides of the field. For example, Adriano of FC Barcelona was playing as left and right back in the season. Messi is considered to be a unique player; this is reflected in our results too with his eighth place in the uniqueness list. We illustrate the most significant feature vectors of these players in the Appendix (Figure~\ref{fig:movement_characteristics_top10}).

\begin{table}[b]
	\scriptsize
	\centering
\begin{tabular}{lrr}
\toprule
  Player &  Uniqueness &  \#movements   \\
\midrule                                                                            
Adriano Correia	& 1.246  &  2067         \\
Martin Montoya	& 1.021 &  1485          \\
Franco Vazquez	& 0.978  &  659          \\
Daniel Larsson	& 0.974 &  738           \\
Oier Sanjurjo Mate	& 0.921 &  2209      \\
Juan Torres Ruiz	& 0.892  &  743      \\
Sergio Ramos	& 0.876 &  2957          \\
Lionel Messi	& 0.860 &  3809          \\
Ruben Garcia Santos	& 0.848 &  1148      \\
Enrique De Lucas	& 0.842 &  611       \\
\bottomrule
\end{tabular}
\caption{The ten most unique players of the competition. The results reveal that it is hard to replace players who are able to play in multiple positions (\eg, different sides of the field).}
\label{tab:uniqueness}
\end{table}

\subsection{Consistency}
\begin{figure}[tb]
\centering
\includegraphics[width=8cm]{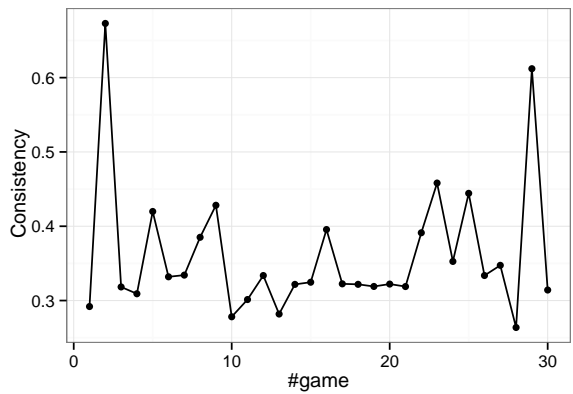}
\caption{The consistency of Xavi throughout the season. His movements were similar and consistent for the majority of the season, the two outliers are games where he was a substitute.}
\label{fig:xavi_consistency}
\end{figure}

\begin{figure}[tb]
\centering
\subfigure[Game 10]{\adjincludegraphics[clip=true, trim=1cm 2.5cm 1cm 3cm,width=5.4cm]{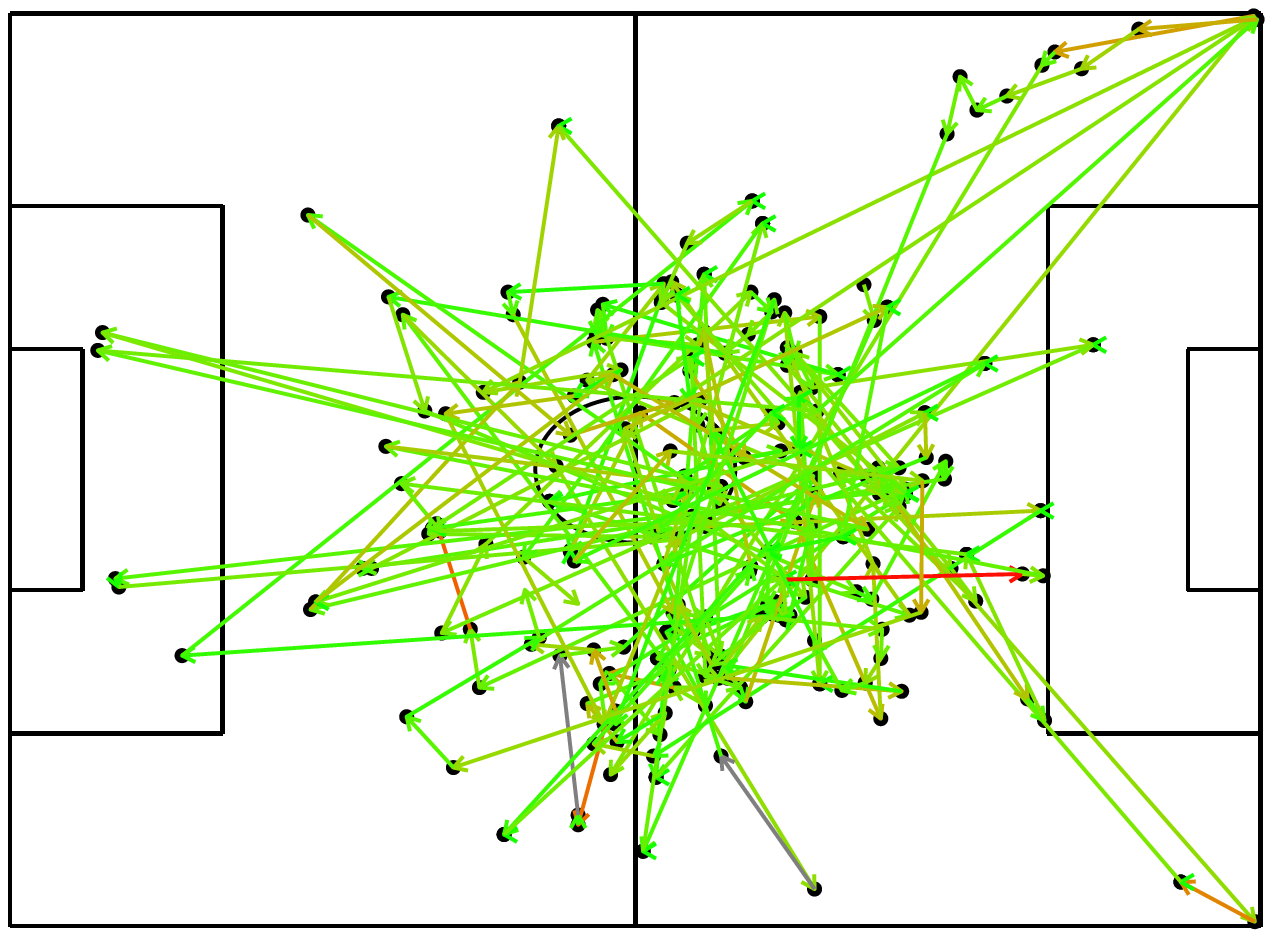}}
\subfigure[Game 12]{\adjincludegraphics[clip=true, trim=1cm 2.5cm 1cm 3cm,width=5.4cm]{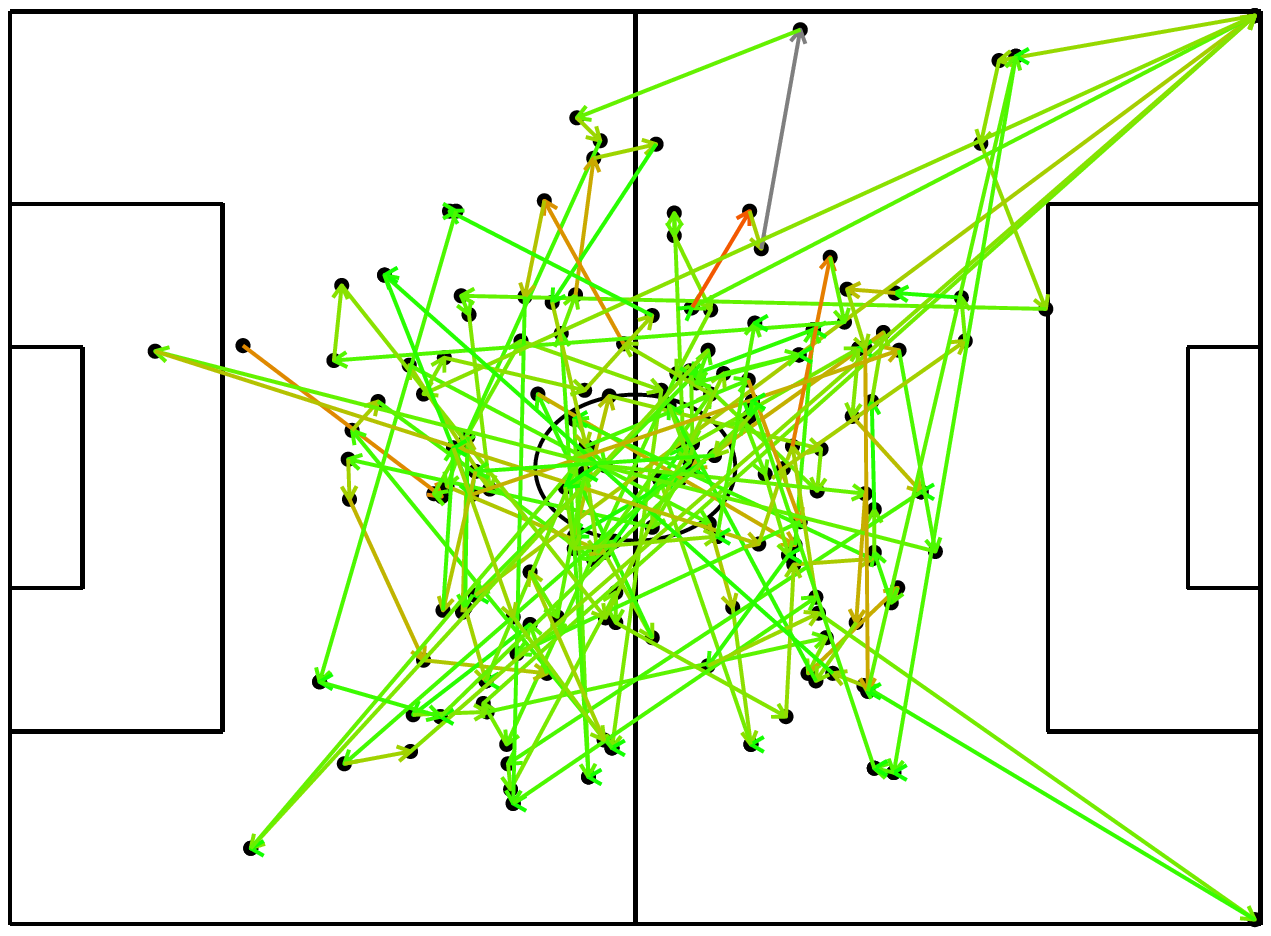}}
\subfigure[Game 23]{\adjincludegraphics[clip=true, trim=1cm 2.5cm 1cm 3cm,width=5.4cm]{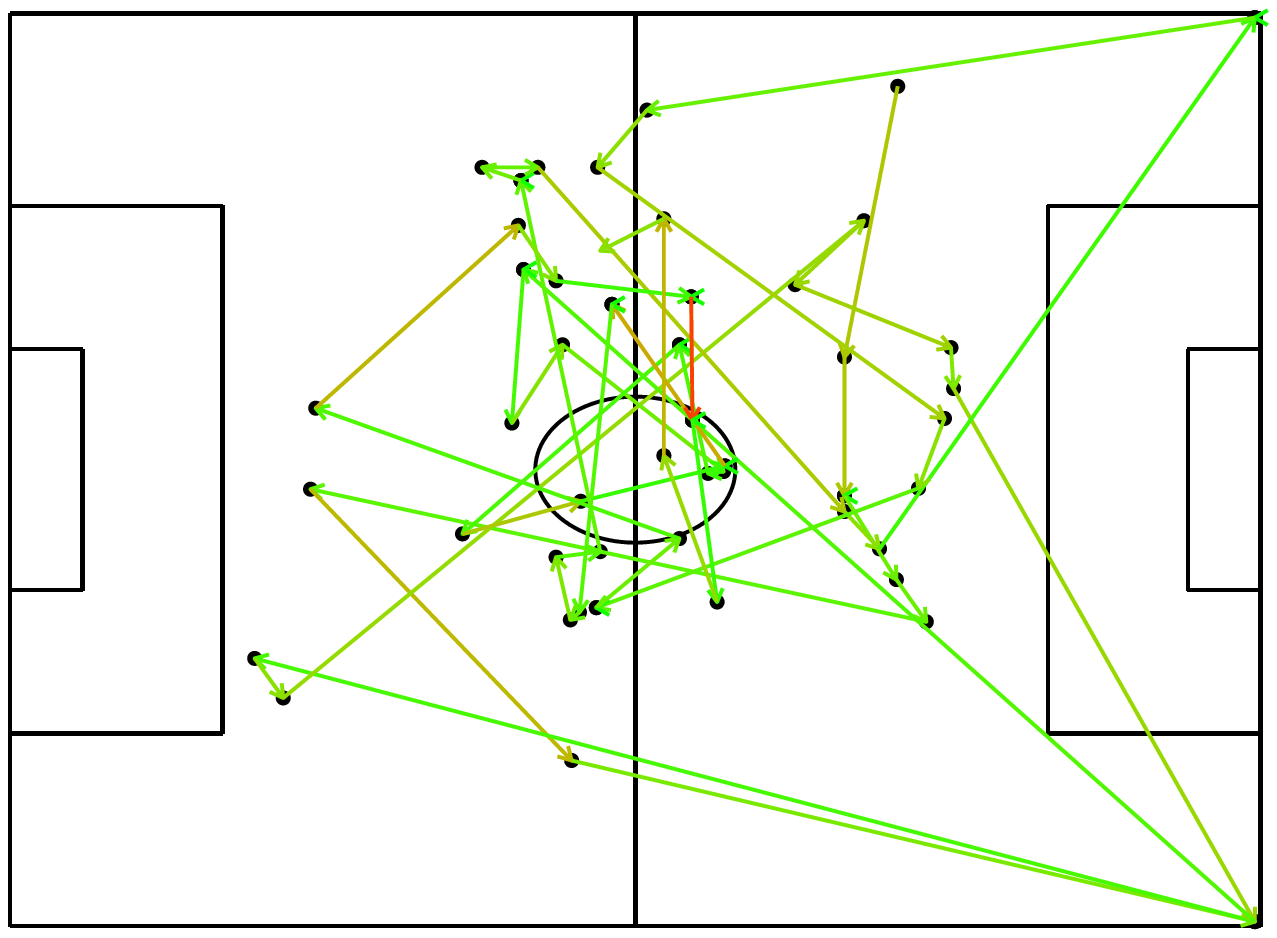}}
\caption{All the movement vectors of Xavi in case of three distinct games. The trends of the movements are similar regardless of the game.}
\label{fig:xavi_details}
\end{figure}

We take one step further and next focus on the consistency of the movements. The managers of the teams prefer players who are consistent. If the performance is consistent, the player will deliver the expected movements. On the other hand, it is hard to count on a player whose movements have high fluctuation. Before analyzing the trends in the league, we first present an example in Figure~\ref{fig:xavi_consistency}, where we show the consistency of Xavi's movements across the season. The horizontal axis denotes the identifier of the game he was involved in, while the vertical axis represents the game-wise consistency. It is remarkable  how consistent Xavi is across the majority of the season. The two outliers at the beginning and at the end of the season are games where he was partially on the bench. We show the movement vectors of game 10, 12, and 23 in Figure~\ref{fig:xavi_details}. The detailed figures highlight similar trajectories. Xavi had numerous lateral movement in the middle of the field, some high intensity movements towards the box of the opponent, and he was responsible for taking the corner kicks of his team. This was Xavi's role in the game against Real Madrid as well, as shown earlier in Figure~\ref{fig:movement_vectors_xavi}. In the third game (\ie, game 23), Xavi moved similarly, however, he was not on the pitch for the entire game, resulting in a slightly elevated consistency value.

\begin{figure}[tb]
\centering
\includegraphics[width=12cm]{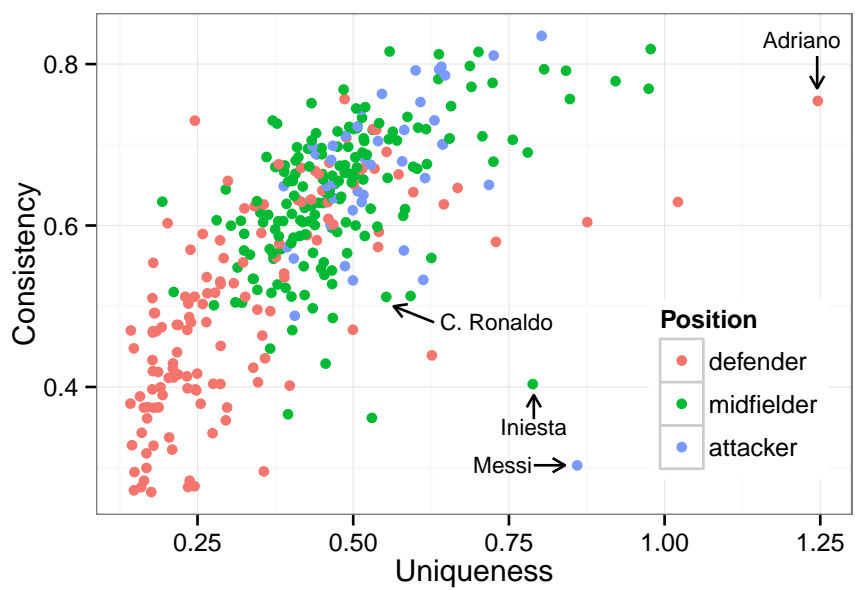}
\caption{The uniqueness and the consistency of the players in the league. The color of the dots denote the position of the players. The outliers of the positions are Adriano, Iniesta, and Messi, respectively.}
\label{fig:uniqueness}
\end{figure}

Finally, we analyze the players of the league in terms of their uniqueness and consistency (Figure~\ref{fig:uniqueness}). Defenders are consistent in general, but they are not having too much distinctness. There is no clear difference between midfielders and attackers. There is a clear relation between the two properties: higher uniqueness comes at a price of consistency. The dataset contains three outliers, oddly all of them are players of FC Barcelona. In terms of the defenders, Adriano has a extremely unique behavior, \ie, high distance from the most similar players, however, his performance is not consistent (consistency of 0.75). Iniesta does movements with high consistency (\ie, value of 0.40) and his movements are fairly unique (0.78). Messi is the outlier of the attackers, with high uniqueness and high consistency (0.86 and 0.30, respectively). Messi's movement profile is in a shocking contrast with Cristiano Ronaldo, who seems to be just an average player in terms of the uniqueness and the consistency of his movements (0.55 and 0.51, respectively).

\subsection{Creating chances}
We are able to extract all in-game movements of the players using our methodology. Hence, we may focus on the movements of the players related to specific events of the game. Potentially, this opens up a new line of research related to the analysis of movements creating scoring opportunities at scale. Here we only highlight the potential of this area by showing two examples. 

\begin{figure}[tb]
\centering
\subfigure[Messi's movements before taking a shot (colors denote the speed of the movements)]{\adjincludegraphics[clip=true, trim=1cm 2.5cm 1cm 3cm,width=7.5cm]{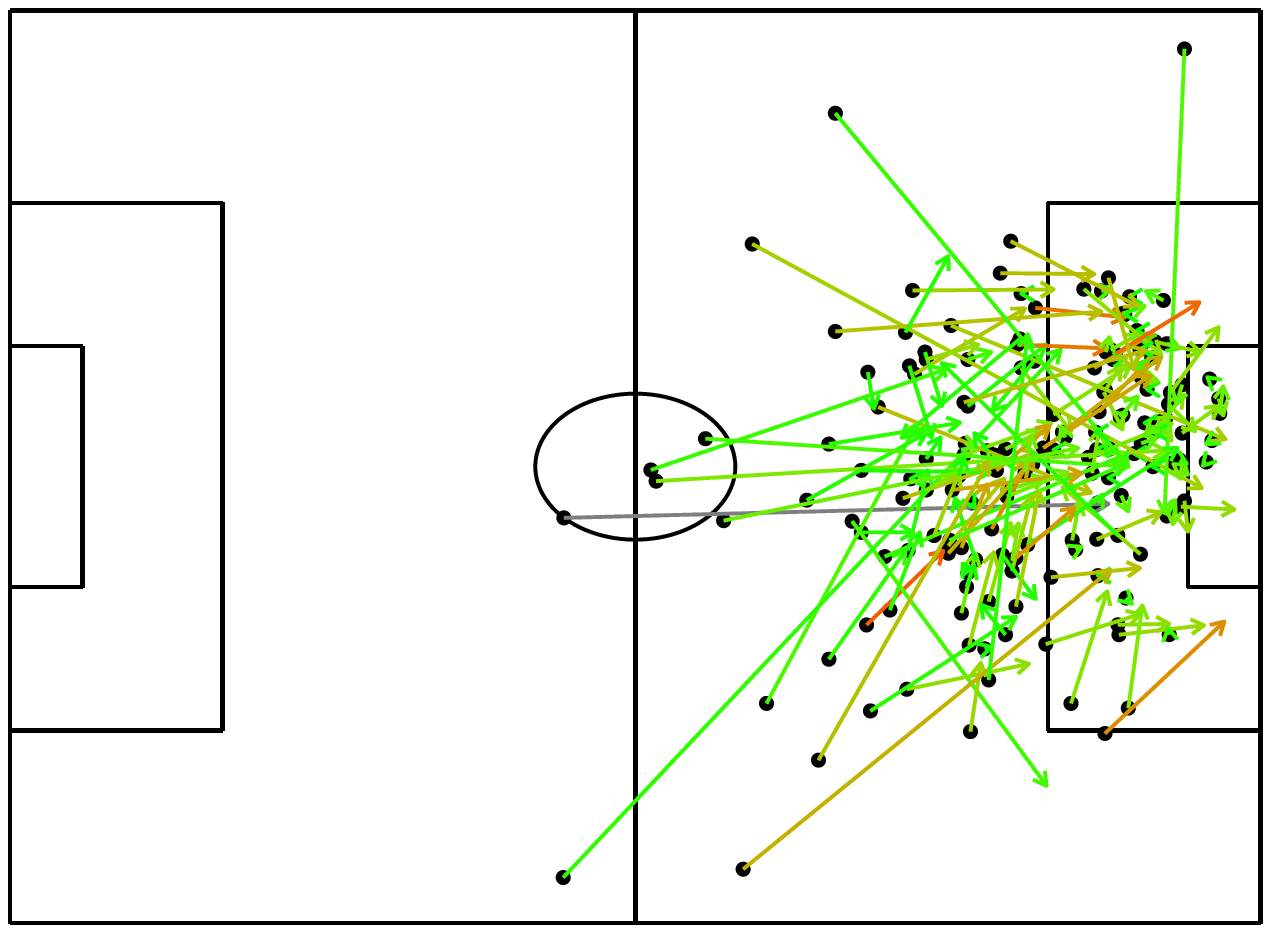}\label{fig:pre_shot_movements_player}}
\subfigure[Movement of a team precluding a shot (colors denote different players of the team)]{\adjincludegraphics[width=7.5cm,clip=true, trim=1cm 2.5cm 1cm 	3cm]{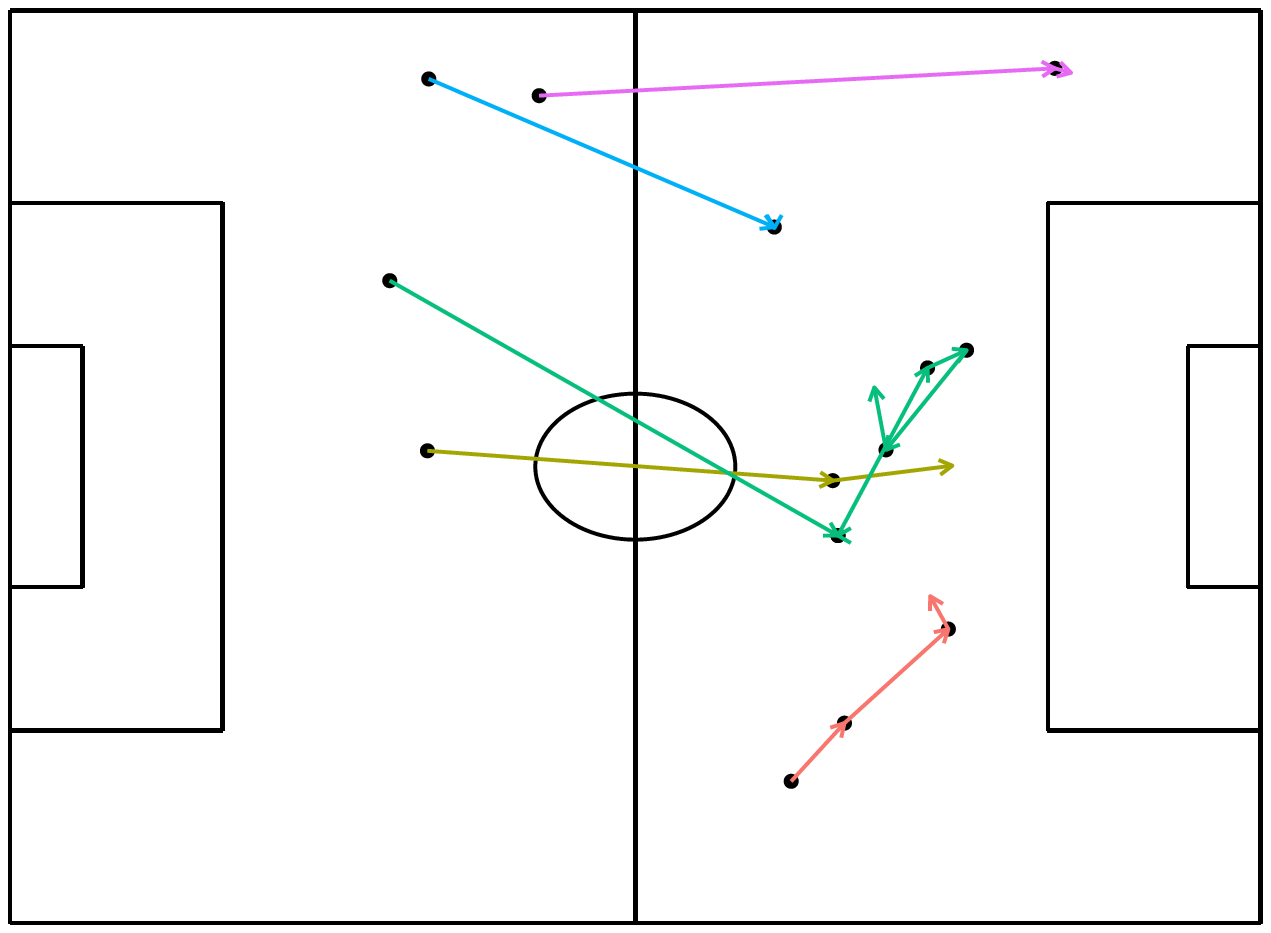}\label{fig:pre_shot_movements_team}}
\caption{Additional application of the methodology: insights on creating chances}
\label{fig:pre_shot_movements}
\end{figure}

Figure~\ref{fig:pre_shot_movements_player} reveals a player's movements before taking a shot. These are the areas from where this player---namely Messi---is able to create chances. There are three main routes to arrive into the final third of the pitch and shoot: running straight on the left side of the box, moving diagonally towards the edge of the box from the middle of the field, and starting from the right side taking a diagonal route towards the box. We note that a notable portion of these movements are high intensity runs.

Most of the time chances are not created by a single player, rather are a result of a series of carefully orchestrated movements of the whole team. We are able to capture this as well with our method as shown in Figure~\ref{fig:pre_shot_movements_team}. The example presents the movements of a team up to 20 seconds earlier than taking a shot. The different colors represent different players of the team. The figure shows intense movements of five players, all of them moving straight towards the opponent's box. This case the goal of the players was pressing the defensive line of the opponent rather than opening up spaces.

These insights are just preliminary steps for a complete, on scale understanding of the relation of player movements and scoring opportunities. A thorough analysis of these phenomena is a future work.

\section{Conclusion}
We analyzed quantitatively the in-game movements of soccer players throughout an entire season. Our methodology reveals detailed insights on the in-game strategy of the teams and on the role and performance of the players within a team. We identify similarities among players and potential candidates for replacing a given player. The results provide valuable inputs for player and opponent scouting by revealing and quantifying the movements of the players to an extent never seen before: both in terms of the number of covered teams and players.


\bibliographystyle{plainnat}
\bibliography{sloan_refs}

\newpage

\section*{Appendix}

\begin{figure}[h]
\centering
\subfigure[Adriano Correia]{\adjincludegraphics[clip=true, trim=1cm 2.5cm 1cm 3cm,width=4.5cm]{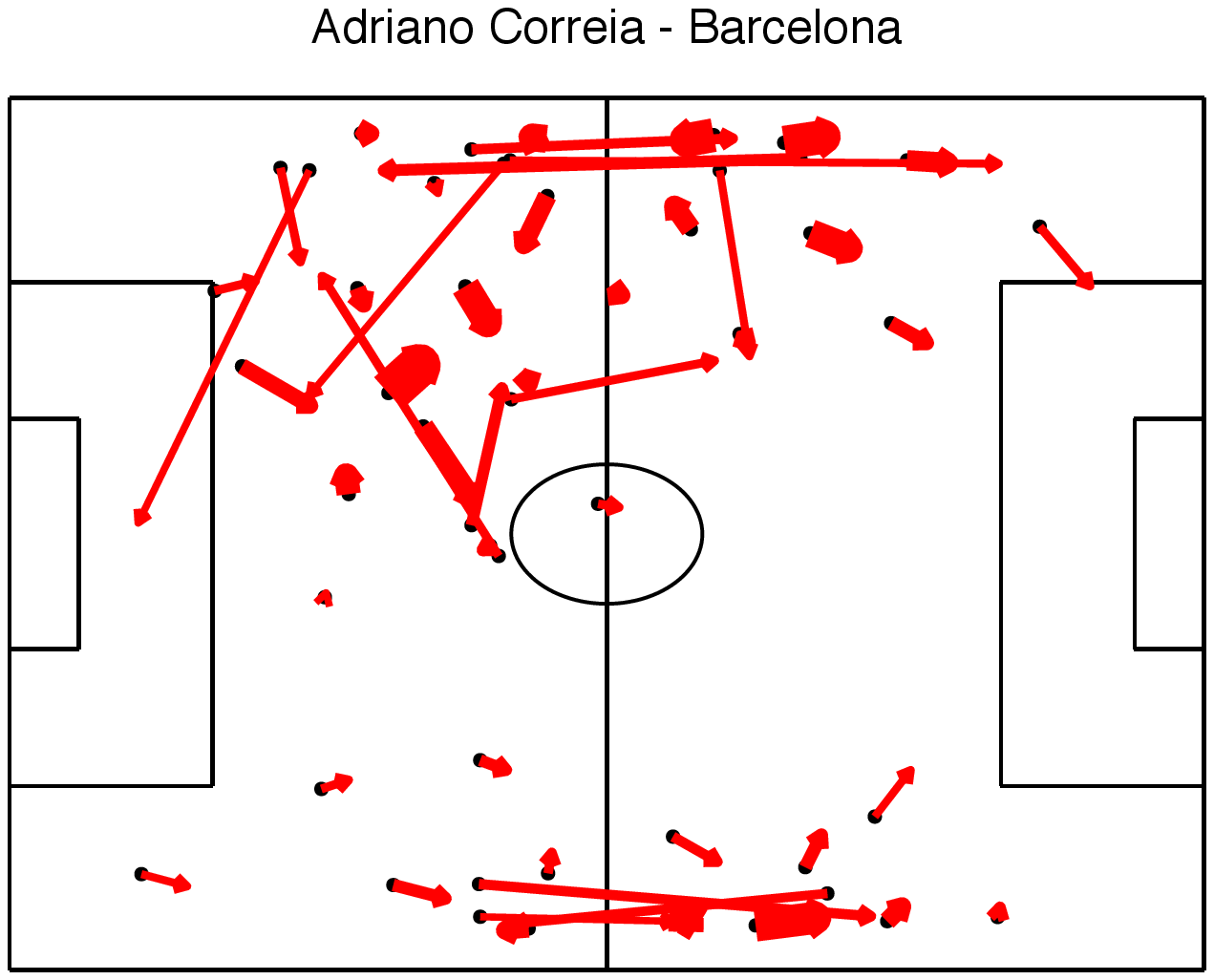}}
\subfigure[Martin Montoya]{\adjincludegraphics[clip=true, trim=1cm 2.5cm 1cm 3cm,width=4.5cm]{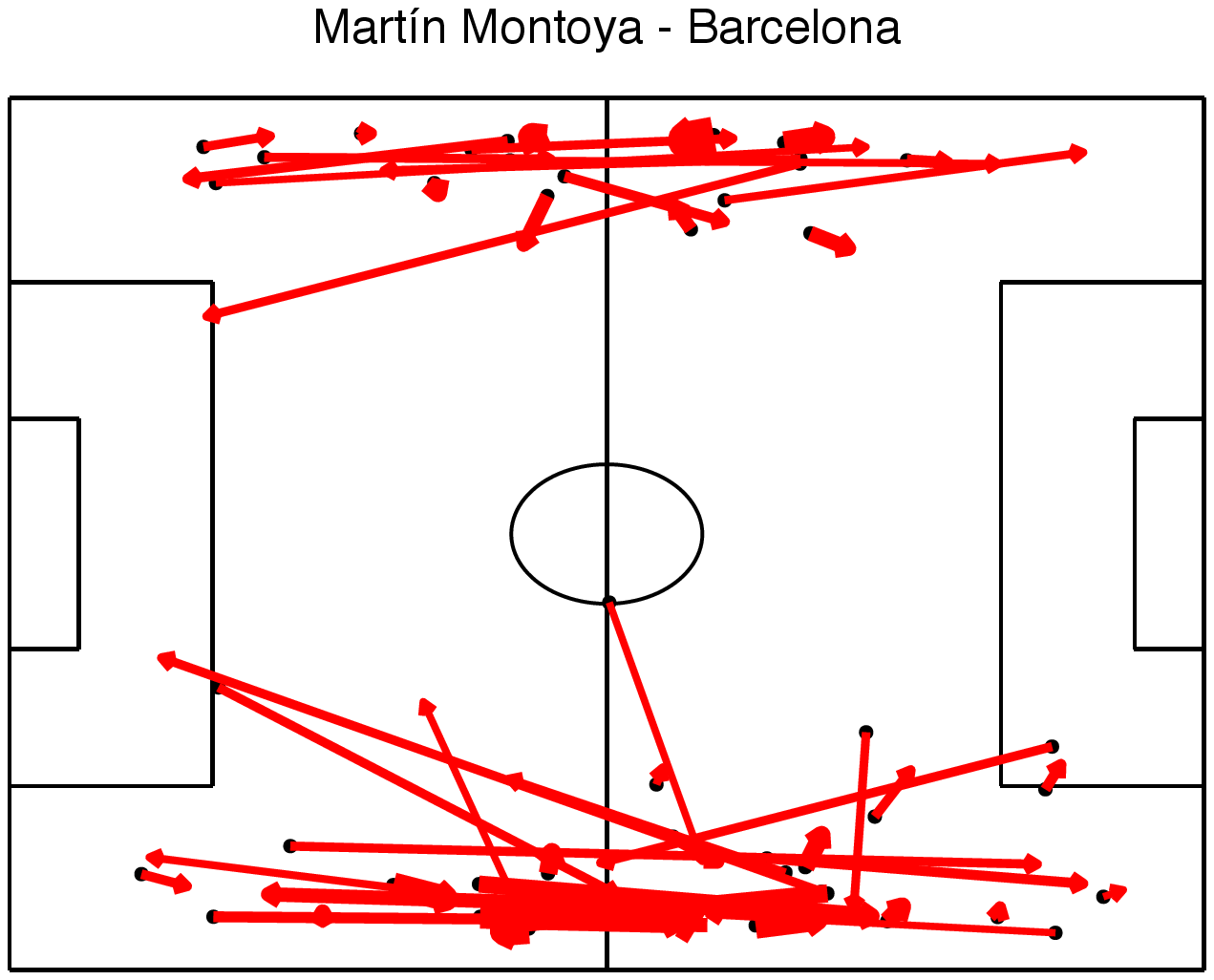}}
\subfigure[Franco Vazquez]{\adjincludegraphics[clip=true, trim=1cm 2.5cm 1cm 3cm,width=4.5cm]{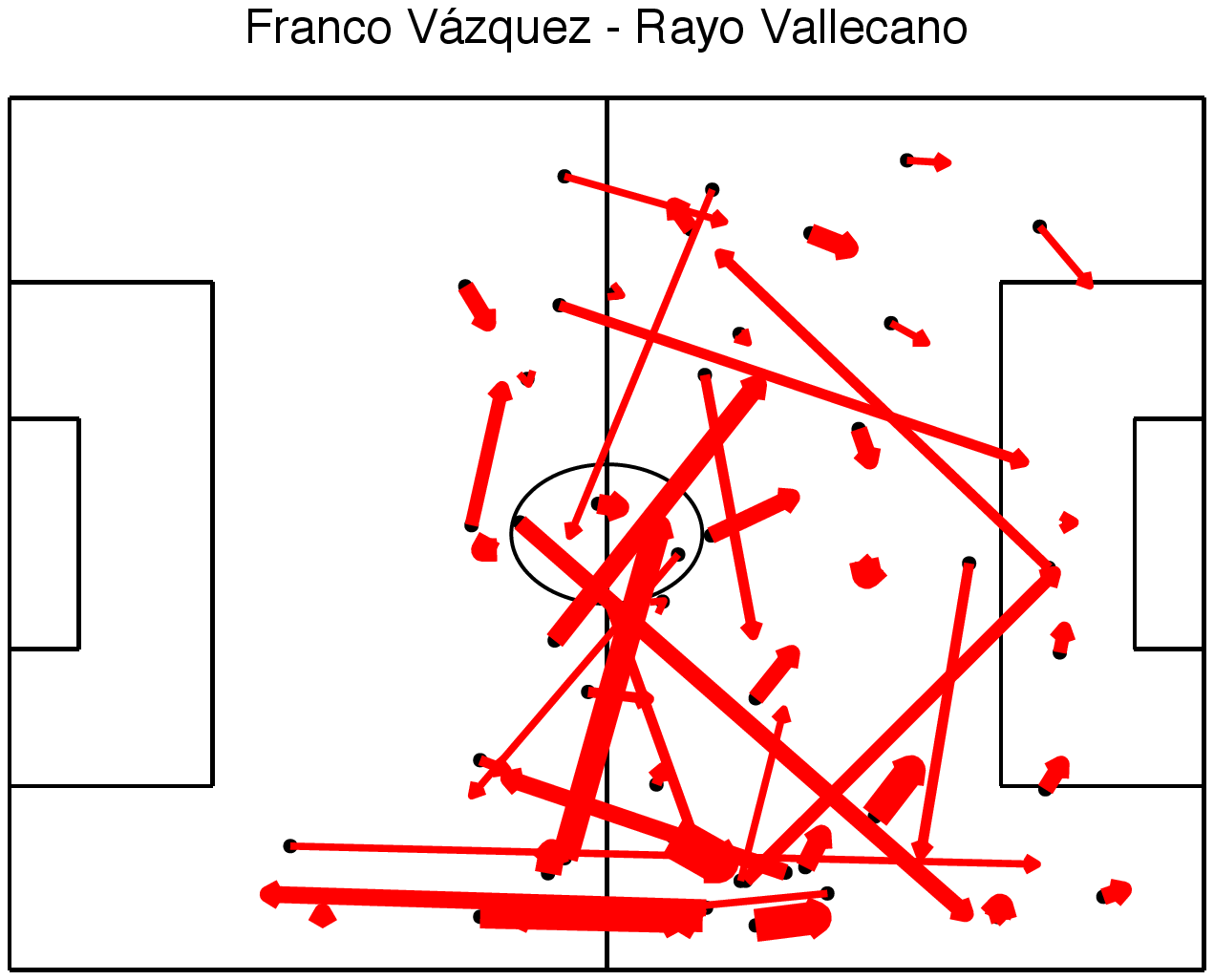}}
\subfigure[Daniel Larsson]{\adjincludegraphics[clip=true, trim=1cm 2.5cm 1cm 3cm,width=4.5cm]{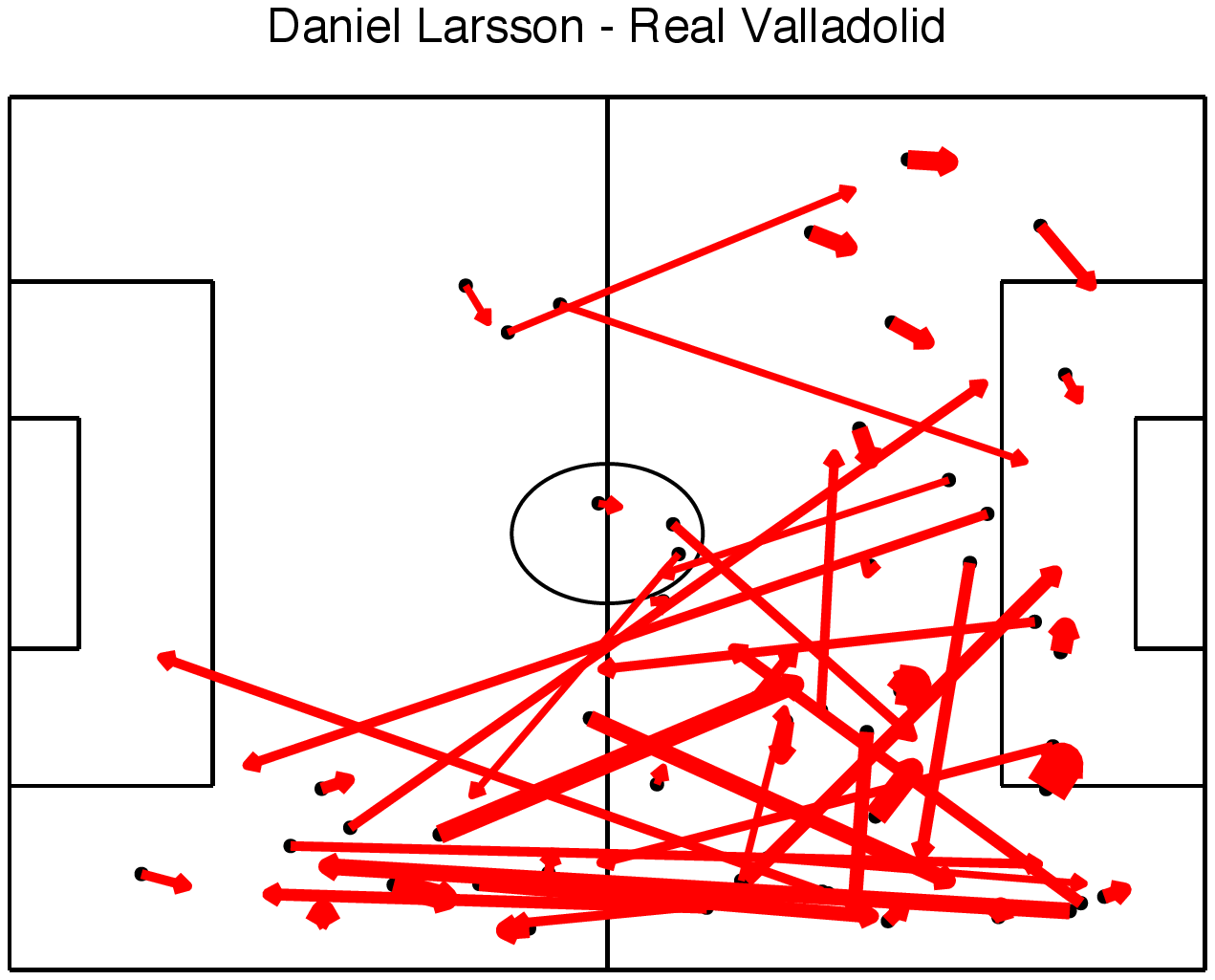}}
\subfigure[Oier Sanjurjo Mate]{\adjincludegraphics[clip=true, trim=1cm 2.5cm 1cm 3cm,width=4.5cm]{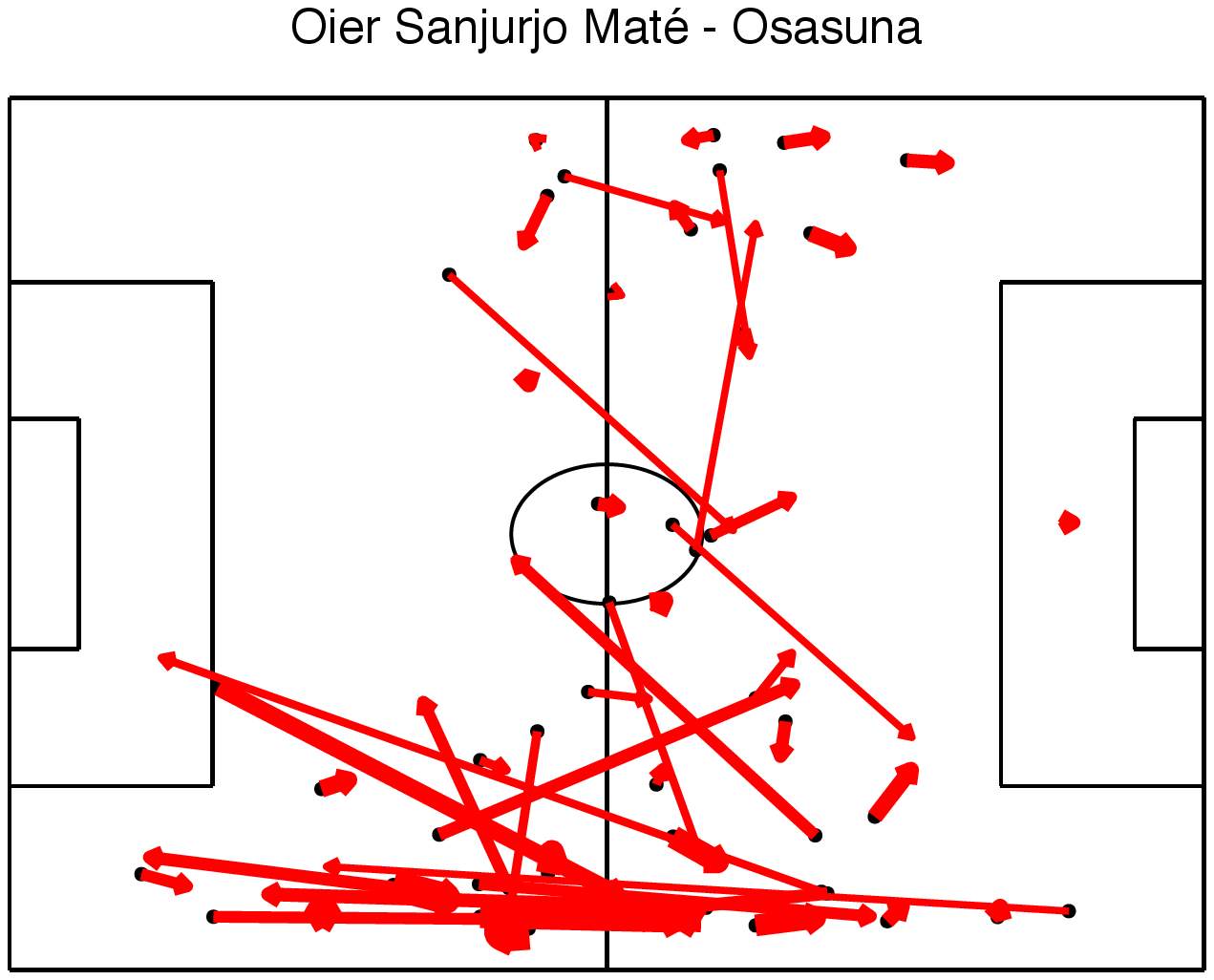}}
\subfigure[Juan Torres Ruiz]{\adjincludegraphics[clip=true, trim=1cm 2.5cm 1cm 3cm,width=4.5cm]{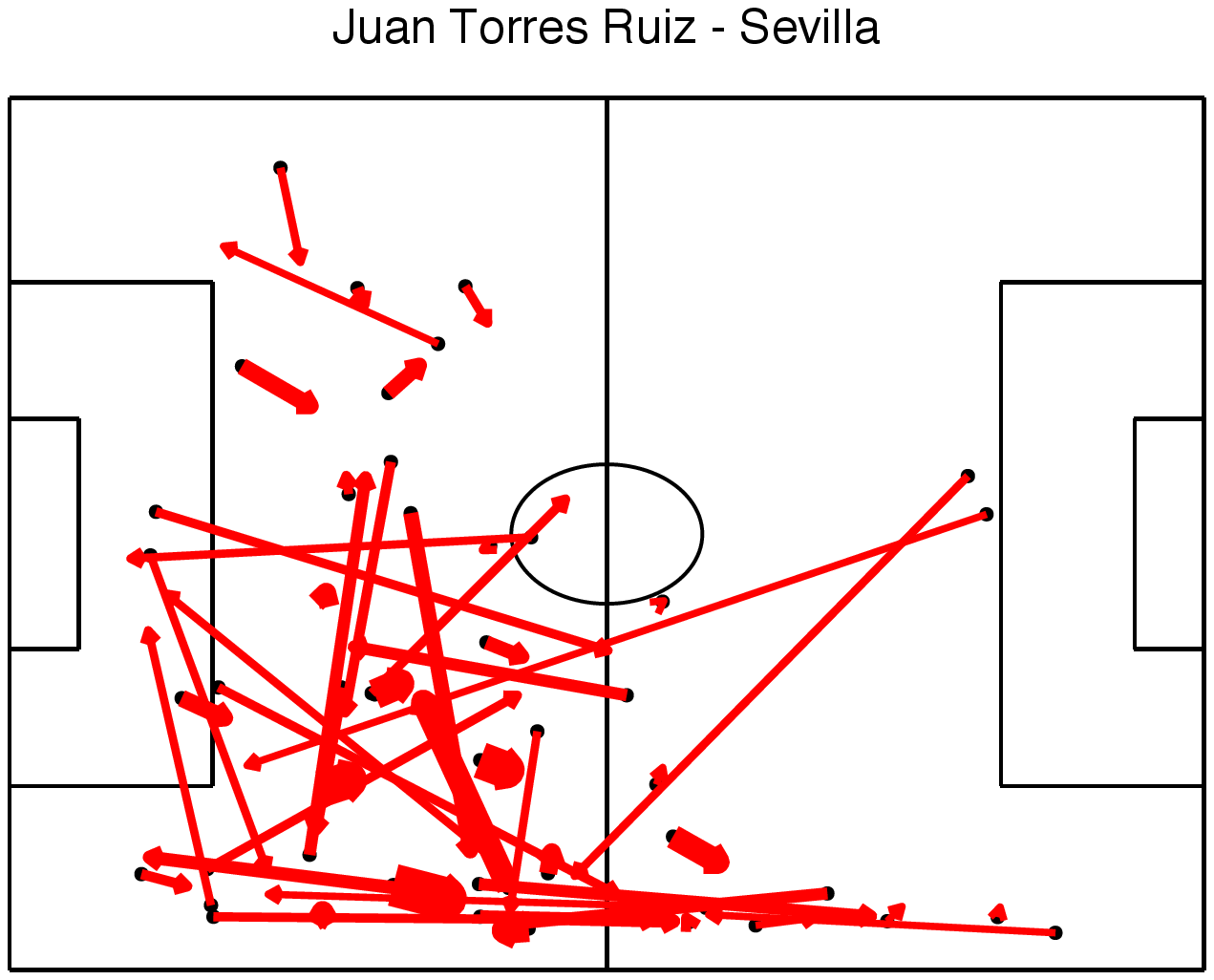}}
\subfigure[Sergio Ramos]{\adjincludegraphics[clip=true, trim=1cm 2.5cm 1cm 3cm,width=4.5cm]{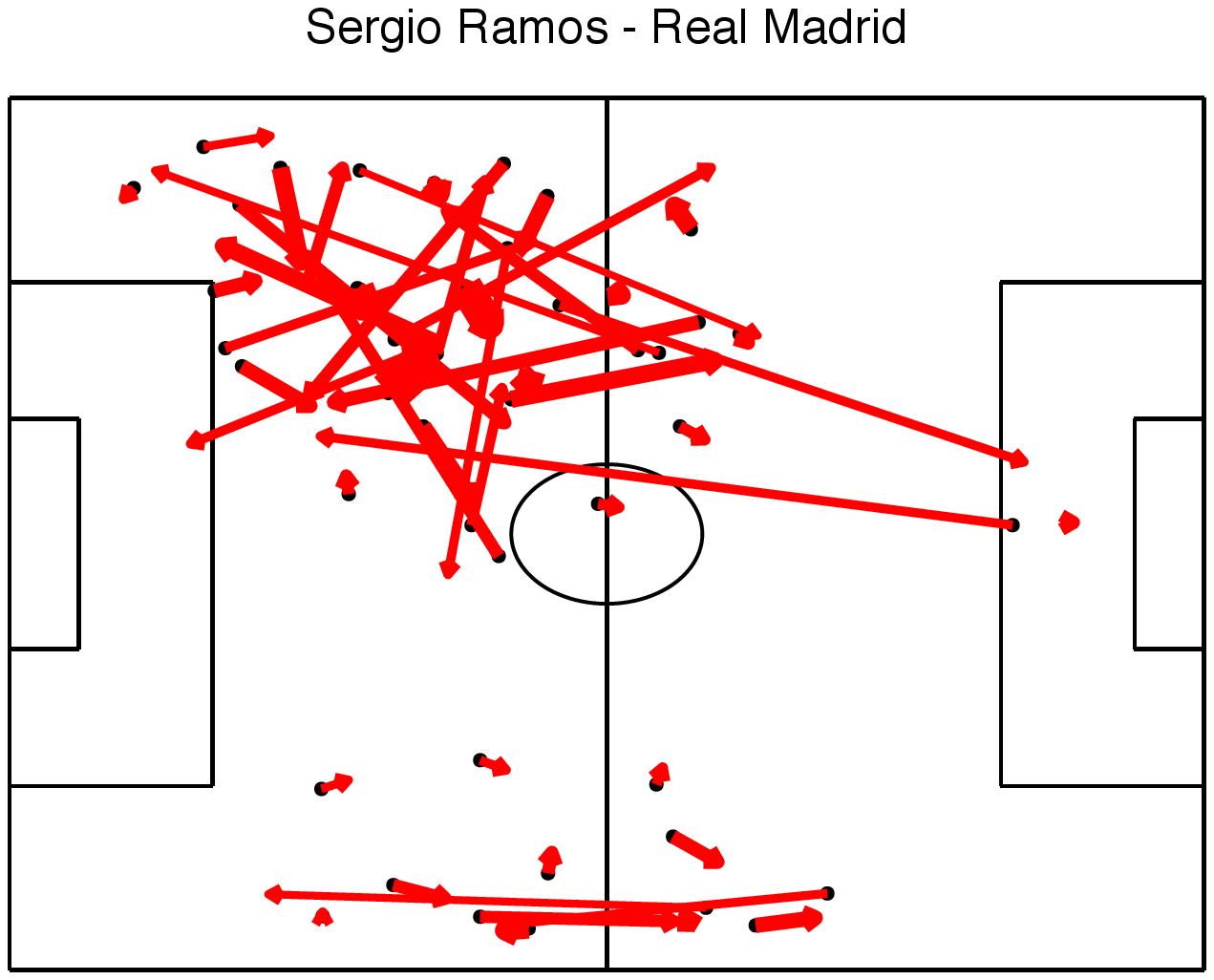}}
\subfigure[Lionel Messi]{\adjincludegraphics[clip=true, trim=1cm 2.5cm 1cm 3cm,width=4.5cm]{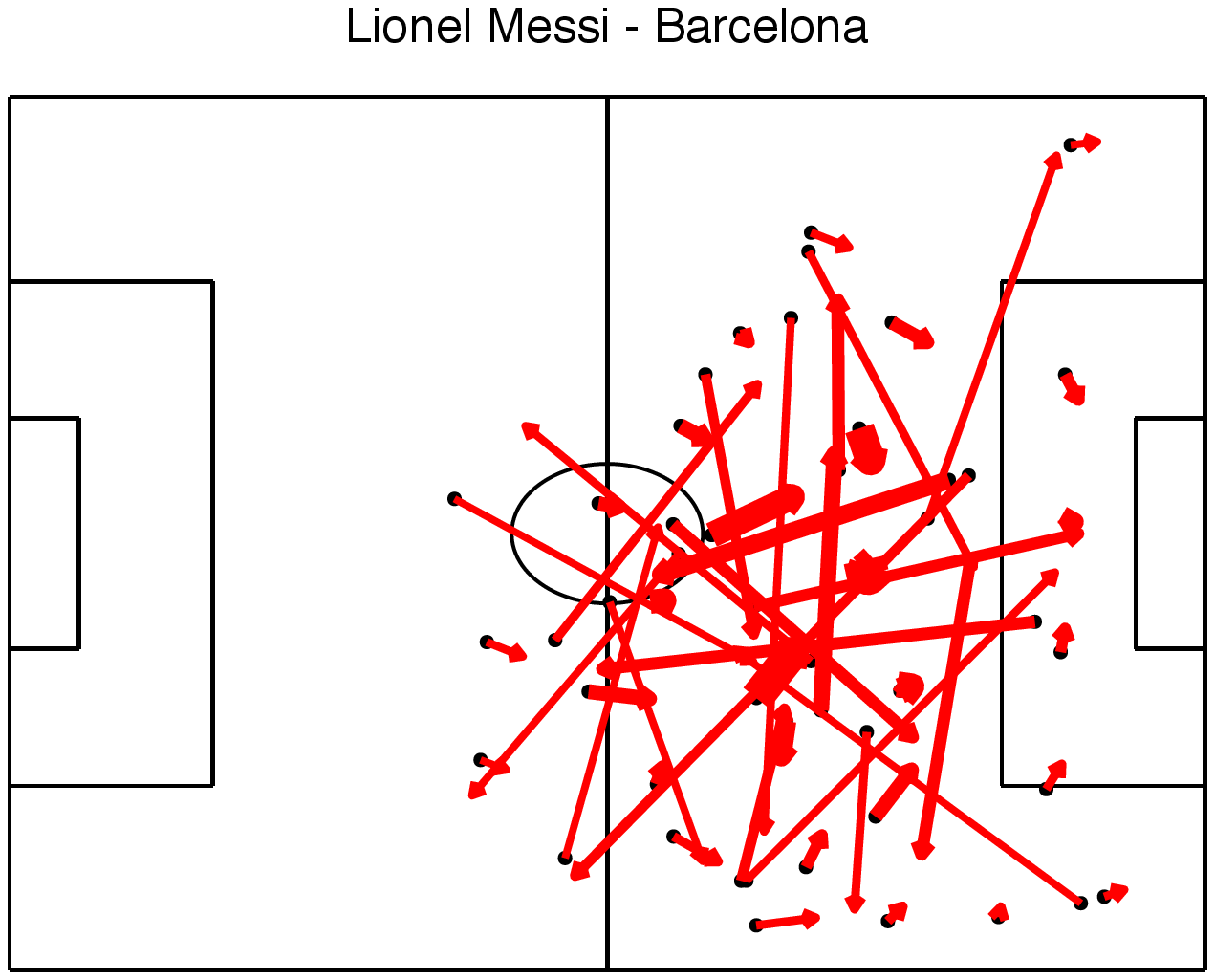}}
\subfigure[Ruben Garcia Santos]{\adjincludegraphics[clip=true, trim=1cm 2.5cm 1cm 3cm,width=4.5cm]{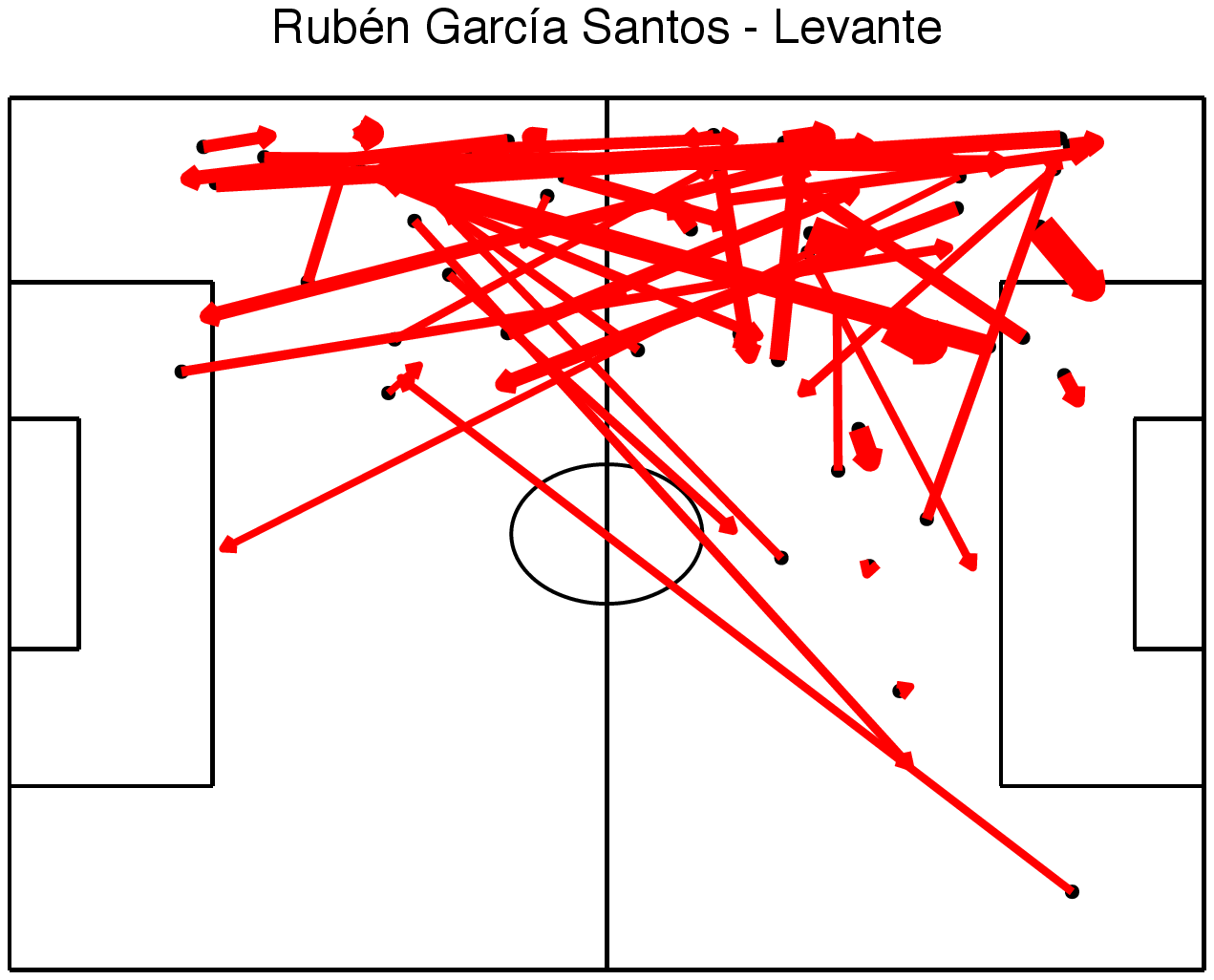}}
\subfigure[Enrique De Lucas]{\adjincludegraphics[clip=true, trim=1cm 2.5cm 1cm 3cm,width=4.5cm]{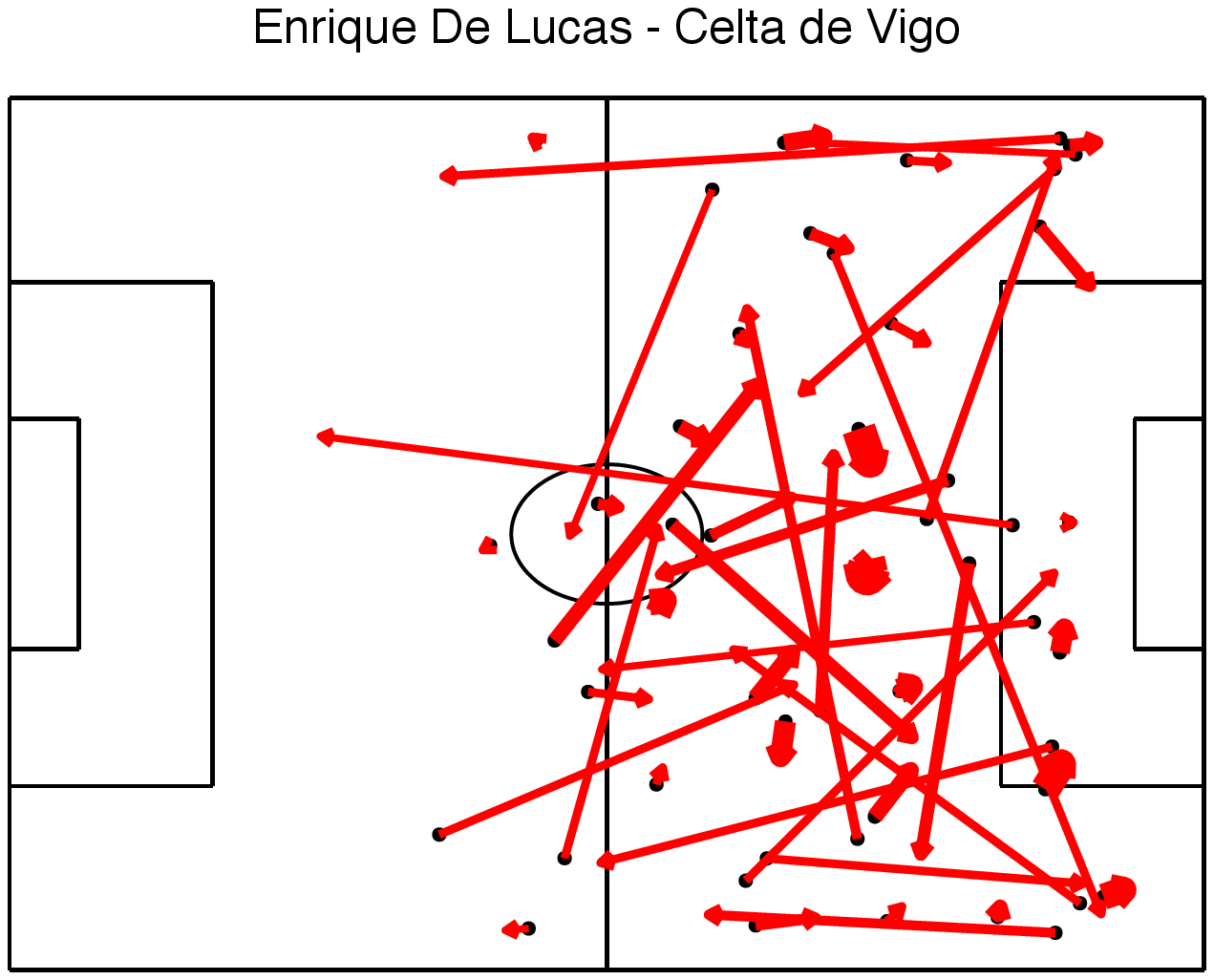}}
\caption{The movement characteristics of the ten most unique players}
\label{fig:movement_characteristics_top10}
\end{figure}


\end{document}